\documentclass[usenatbib]{mn2e}
\usepackage{amsfonts}
\usepackage{txfonts}
\usepackage{mathrsfs}
\usepackage{amssymb}
\usepackage{psfig}
\usepackage{graphicx,times}
\usepackage{natbib}

\footnotesize

\newcommand\nuddd{\ifmmode\stackrel{\bf \,...}{\textstyle \nu}\else$\stackrel{\,...}{\textstyle \nu}$\fi}
\def\lsim{~\rlap{$<$}{\lower 1.0ex\hbox{$\sim$}}}
\def\gsim{~\rlap{$>$}{\lower 1.0ex\hbox{$\sim$}}}

\title{A molecular line study toward massive EGO clumps in the southern sky: Chemical properties}

\author[Naiping Yu, Jun-Jie Wang and Nan Li]{Naiping Yu$^{1,2}$\thanks{E-mail: yunaiping09@mails.gucas.ac.cn},
Jun-Jie Wang$^{1,2}$ \footnotemark[1]\thanks{}\\
$^{1}$National Astronomical Observatories, Chinese Academy of
Sciences, Beijing 100012, China\\
$^{2}$NAOC-TU Joint Center for Astrophysics, Lhasa 850000, China}

\begin{document}
\maketitle \pagestyle{plain}

\begin{abstract}
We present a molecular line study toward 31 extended green object
(EGO) clumps in the southern sky using data from the MALT90
(Millimeter Astronomy Legacy Team 90 GHz). According to previous
multiwavelength observations, we divide our sample into two groups:
massive young stellar objects (MYSOs) and HII regions. The most
detected lines are N$_2$H$^+$ ($J=1-0$), HCO$^+$ ($J=1-0$), HNC
($J=1-0$), HCN ($J=1-0$), HC$_3$N ($J=10-9$), H$^{13}$CO$^+$
($J=1-0$), C$_2$H ($N=1-0$) and SiO ($J=2-1$), indicating that most
EGOs are indeed associated with dense clumps and recent outflow
activities. The velocity widths of the N$_2$H$^+$ ($J=1-0$),
H$^{13}$CO$^+$ ($J=1-0$), C$_2$H ($N=1-0$) and HC$_3$N ($J=10-9$)
lines are comparable to each other in MYSOs. However, in HII regions
the velocity widths of the N$_2$H$^+$ ($J=1-0$) and C$_2$H ($N=1-0$)
lines tend to be narrower than those of H$^{13}$CO$^+$ ($J=1-0$) and
HC$_3$N ($J=10-9$). Our results seem to support that N$_2$H$^+$ and
C$_2$H emissions mainly come from the gas inside quiescent clumps.
In addition, we also find that the [N$_2$H$^+$]/[H$^{13}$CO$^+$] and
[C$_2$H]/[H$^{13}$CO$^+$] relative abundance ratios decrease from
MYSOs to HII regions. These results suggest depletions of N$_2$H$^+$
and C$_2$H in the late stages of massive star formation, probably
caused by the formation of HII regions inside. N$_2$H$^+$ and C$_2$H
might be used as ``chemical clocks" for massive star formation by
comparing with other molecules such as H$^{13}$CO$^+$ and HC$_3$N.

\end{abstract}

\begin{keywords}
stars: formation - ISM: abundances - ISM: molecules - ratio lines:
ISM - ISM: clouds
\end{keywords}

\begin{table*}
\begin{minipage}{13cm}
 \caption{\label{tab:test}List of our sources.}
 \begin{tabular}{lclclclclclclcl}
  \hline
  \hline
 EGO  & R.A.  & Dec.  & V$_{lsr}^a$ & IRDC? & Type& \\

 name    & (J2000) & (J2000) &(km s$^{-1}$) & & & \\
  \hline
  MYSOs \\
  \hline
G10.34$-$0.14 & 18:09:00.0 & -20:03:35 & 12.4 & Y & MSX Dark \\
G14.63$-$0.58 & 18:19:15.4 & -16:30:07 & 18.9 & Y & MSX Dark \\
G309.38$-$0.13& 13:47:23.9 & -62:18:12 & -50.5& Y & MSX Dark \\
G309.91+0.32& 13:50:53.9 & -61:44:22 & -59.3 & Y & MSX Dark \\
G313.76-0.86& 14:25:01.3 & -61:44:57 & -50.7 & Y & radio-quiet RMS \\
G326.41+0.93& 15:41:59.4 & -53:59:03 & -41.3 & Y & MSX Dark \\
G326.48+0.70& 15:43:17.5 & -54:07:11 & -41.0 & Y & radio-quiet RMS \\
G327.30-0.58& 15:53:11.2 & -54:36:48 & -45.6 & Y & MSX Dark \\
G329.07-0.31& 16:01:11.7 & -53:16:00 & -42.2 & Y & radio-quiet RMS \\
G331.51-0.34& 16:13:11.7 & -51:39:12 & -48.5 & Y & MSX Dark \\
G331.71+0.58& 16:10:06.3 & -50:50:29 & -66.6 & Y & MSX Dark \\
G331.71+0.60& 16:10:01.9 & -50:49:33 & -67.6 & Y & radio-quiet RMS \\
G332.35-0.44& 16:17:31.4 & -51:08:22 & -48.1 & Y & MSX Dark \\
G332.60-0.17& 16:17:29.4 & -50:46:13 & -46.4 & Y & MSX Dark \\
G332.47-0.52& 16:18:26.5 & -51:07:12 & -51.3 & Y & radio-quiet RMS \\
G333.13-0.56& 16:21:36.1 & -50:40:49 & -58.8 & N & MSX Dark \\
G333.32+0.10& 16:19:28.9 & -50:04:40 & -46.2 & Y & radio-quiet RMS \\
G338.92+0.55& 16:40:33.6 & -45:41:44 & -62.7 & N & radio-quiet RMS \\
G340.97-1.02& 16:54:57.3 & -45:09:04 & -24.0 & Y & MSX Dark \\
G343.50+0.03& 16:59:10.7 & -42:31:07 & -29.5 & Y & MSX Dark \\

  \hline
  HII regions \\
  \hline
G12.42+0.50& 18:10:51.1 & -17:55:50 & 18.3 & N & radio-loud RMS \\
G310.15+0.76& 13:51:59.2 & -61:15:37 & -54.7 & N & radio-loud RMS \\
G320.23-0.28& 15:09:52.6 & -58:25:36 & -65.9 & Y & radio-loud RMS \\
G327.40+0.44& 15:49:19.3 & -53:45:10 & -79.6 & Y & radio-loud RMS \\
G330.88-0.37& 16:10:19.9 & -52:06:13 & -62.1 & Y & radio-loud RMS \\
G330.95-0.18& 16:09:52.7 & -51:54:56 & -88.9 & Y & radio-loud RMS \\
G331.13-0.24& 16:10:59.8 & -51:50:19 & -85.8 & N & radio-loud RMS \\
G337.40-0.40& 16:38:50.4 & -47:28:04 & -40.4 & Y & radio-loud RMS \\
G339.58-0.13& 16:45:59.5 & -45:38:44 & -33.7 & Y & radio-loud RMS \\
G340.05-0.25& 16:48:14.7 & -45:21:52 & -52.9 & Y & radio-loud RMS \\
G345.00-0.22& 17:05:11.2 & -41:29:03 & -27.7 & Y & radio-loud RMS \\
 \hline
 \end{tabular}
 \label{tb:rotn}

 a: Velocity from the N$_2$H$^+$ ($J=1-0$, $F_1=2-1$, $F=3-2$)
 transition.
\end{minipage}
\end{table*}

\begin{table*}
\begin{minipage}{13cm}
 \caption{\label{tab:test}Velocity Widths of these detected lines.}
 \begin{tabular}{lclclclclclclcl}
  \hline
  \hline
 EGO  & N$_2$H$^+$  & H$^{13}$CO$^+$ & HCO$^+$ & HNC  & C$_2$H & HC$_3$N & SiO& \\
 name & (km s$^{-1}$) & (km s$^{-1}$) & (km s$^{-1}$)& (km
 s$^{-1}$) & (km s$^{-1}$) & (km s$^{-1}$) & (km s$^{-1}$) & \\
  \hline
  MYSOs \\
  \hline
G10.34$-$0.14 & 3.1(0.1) & 2.2(0.3) & 3.7(0.1) & 4.2(0.1) & 3.5(0.3) & 3.4(0.2) & - - - \\
G14.63$-$0.58 & 3.0(0.1) & 2.9(0.3) & R$^a$        & B$^b$        & 3.2(0.2)    & 2.5(0.1) & - - - \\
G309.38$-$0.13& 3.2(0.1) & - - -    & R        & 5.4(0.2) & - - -  & 4.0(0.2) & - - - \\
G309.91+0.32& 3.1(0.1) & - - - & 5.3(0.3)        & 4.3(0.2)        & - - -    & 2.4(0.3) & - - - \\
G313.76-0.86& 3.8(0.1) & 3.3(0.3) & 5.6(0.1)   & 4.8(0.1)     & 4.7(0.4)   & 3.8(0.3) & - - - \\
G326.41+0.93& 3.0(0.1) & - - - & R        & 4.1(0.3)        & - - -    & 6.7(1.1) & - - - \\
G326.48+0.70& 3.6(0.1) & 2.4(0.2) & 5.1(0.1)     & 4.3(0.1)     & 3.6(0.3)    & 3.3(0.2) & 7.2(0.9) \\
G327.30-0.58& 4.1(0.1) & 4.9(0.2) & C$^c$        & C        & 6.2(0.2)    & 5.5(0.1) & 7.1(0.3) \\
G329.07-0.31& 5.4(0.1) & - - - & 4.2(0.2)      & 4.3(0.1)     & 3.5(0.4)    & 5.2(0.4) & - - - \\
G331.51-0.34& 2.2(0.1) & - - - & 2.4(0.1)     & 2.1(0.1)     & - - -    & - - - & - - - \\
G331.71+0.58& 5.0(0.2) & 3.8(0.5) & B        & R        & - - -    & 4.5(0.4) & 12.6(1.4) \\
G331.71+0.60& 3.9(0.1) & 4.3(0.5) & B        & B        & 3.8(0.4)    & 2.8(0.2) & 5.1(0.5) \\
G332.35-0.44& 2.5(0.1) & - - - & 2.8(0.1)     & 2.2(0.1)     & - - -    & - - - & - - - \\
G332.47-0.52& 3.5(0.1) & 2.8(0.3) & R        & 5.2(0.2)     & 3.6(0.2)    & 3.4(0.2) & - - - \\
G332.60-0.17& 3.2(0.1) & - - - & B        & 3.8(0.1)        & - - -    & - - - & - - - \\
G333.13-0.56& 3.5(0.1) & - - - & 4.5(0.2)     & 4.0(0.2)     & - - -    & 5.2(0.4) & 8.7(1.0) \\
G333.32+0.10& 3.6(0.1) & - - - & R        & 4.5(0.1)     & 3.3(0.5)    & 3.6(0.3) & - - - \\
G338.92+0.55& 7.2(0.2) & 6.9(0.5) & B        & B        & 7.1(0.5)    & 6.5(0.3) & 7.9(0.6) \\
G340.97-1.02& 4.3(0.1) & 4.2(0.3) & B        & B        & 4.1(0.2)    & 4.6(0.2) & 7.8(0.4) \\
G343.50+0.03& 2.9(0.1) & 2.7(0.4) & 3.4(0.1)      & 2.9(0.1)      & - - -    & - - - & - - - \\

  \hline
  HII regions \\
  \hline
G12.42+0.50& 3.2(0.1) & 2.6(0.1) & B        & B        & 3.1(0.1)    & 2.9(0.2) & - - - \\
G310.15+0.76& 3.5(0.2) & 2.2(0.2) & 7.9(0.2)     & 4.4(0.2)      & 2.7(0.2)    & 4.4(0.5) & - - - \\
G320.23-0.28& 3.8(0.1) & 3.6(0.5) & 6.4(0.2)     & 4.9(0.1)     & 3.6(0.2)   & 2.9(0.2) & - - - \\
G327.40+0.44& 4.1(0.2) & 5.0(0.7) & B        & B        & 4.0(0.4)    & 6.2(0.5) & 7.2(0.6) \\
G330.88-0.37& 4.2(0.3) & 4.4(0.4) & 6.3(0.1)  & 6.6(0.1)    & 4.7(0.2)    & 4.9(0.2) & - - - \\
G330.95-0.18& 11.1(0.6) & 8.2(0.8) & B        & 8.6(0.3)    & 7.1(0.5)    & 9.9(0.6) & 8.9(0.8) \\
G331.13-0.24& 5.7(0.2) & - - - & B        & B        & - - -    & 6.8(0.3) & - - - \\
G337.40-0.40& 3.6(0.1) & 4.1(0.2) & R        & B        &4.0(0.2)   & 4.9(0.2) & 7.6(0.6) \\
G339.58-0.13& 3.5(0.1) & - - - & 6.5(0.3)    & 4.8(0.2)     & - - -    & - - - & - - - \\
G340.05-0.25& 3.8(0.1) & 5.7(0.4) & 7.6(0.2)    & B        & 3.7(0.3)    & 4.8(0.2) & - - - \\
G345.00-0.22& 4.1(0.1) & 5.0(0.5) & 5.5(0.3)   & R        & 6.3(0.6)   & 7.0(0.3) & 16.8(0.8) \\
 \hline
 \end{tabular}
 \label{tb:rotn}
a: R denotes red profile.

b: B denotes blue profile.

c: C denotes the spectra is too complex.
\end{minipage}
\end{table*}

\begin{table*}
\scriptsize
\begin{minipage}{22cm}
 \caption{\label{tab:test}Intensities and integrated intensities of N$_2$H$^+$, HCO$^+$, H$^{13}$CO$^+$ and C$_2$H. }
 \begin{tabular}{lclclclclclclcl}
  \hline
  \hline
 EGO  & \multicolumn{2}{|c|}{N$_2$H$^+$}& \multicolumn{2}{|c|}{N$_2$H$^+$}& \multicolumn{2}{|c|}{HCO$^+$} &  \multicolumn{2}{|c|}{H$^{13}$CO$^+$} & \multicolumn{2}{|c|}{C$_2$H} & \multicolumn{2}{|c|}{C$_2$H}& \\
& \multicolumn{2}{|c|}{group 1} &\multicolumn{2}{|c|}{group 2} & & & & & \multicolumn{2}{|c|}{$F = 1-0$} & \multicolumn{2}{|c|}{$F = 2-1$} \\
 name   & $T_{mb}$ & $\int$$T_{mb}$$dv$ & $T_{mb}$ & $\int$$T_{mb}$$dv$ & $T_{mb}$ & $\int$$T_{mb}$$dv$ & $T_{mb}$ & $\int$$T_{mb}$$dv$ & $T_{mb}$ & $\int$$T_{mb}$$dv$ & $T_{mb}$ & $\int$$T_{mb}$$dv$ \\
 \hline
 & K & K km s$^{-1}$  & K & K km s$^{-1}$  & K & K km s$^{-1}$  & K & K km s$^{-1}$  & K & K km s$^{-1}$  & K & K km s$^{-1}$ \\
  \hline
  MYSOs \\
  \hline
G10.34$-$0.14 & 1.32 & 3.96 (0.36) & 5.25 & 17.54 (0.37) & 4.17 & 16.39 (0.27) & 0.85 & 1.96 (0.22) & 0.70 & 2.78 (0.26) & 1.45 & 5.55 (0.29) \\
G14.63$-$0.58 & 1.85 & 3.96 (0.18) & 5.02 & 16.23 (0.19) & 2.45 & 10.61 (0.36) & 0.89 & 2.72 (0.22) & 0.95 & 2.98 (0.21) & 1.34 & 4.66 (0.22)\\
G309.38$-$0.13& 0.80 & 2.19 (0.24) & 2.58 & 08.93 (0.26) & 2.52 & 07.35 (0.39) & - - - & - - - & - - - &- - - &- - - &- - - \\
G309.91+0.32& 1.61 & 2.86 (0.19) & 3.62 & 11.96 (0.24) & 1.81 & 10.18 (0.38) & - - - &- - - & - - - &- - - &- - - &- - - \\
G313.76-0.86& 1.43 & 3.69 (0.24) & 4.70 & 18.85 (0.31) & 6.84 & 40.49 (0.40) & 1.31 & 4.56 (0.30) & 0.61 & 2.78 (0.33) & 1.11 & 5.40 (0.35)\\
G326.41+0.93& 1.24 & 2.79 (0.22) & 5.30 & 17.04 (0.23) & 2.10 & 08.18 (0.68) & - - - &- - - & - - - &- - - &- - - &- - -  \\
G326.48+0.70& 1.62 & 7.29 (0.45) & 6.08 & 23.35 (0.47) & 6.19 & 33.77 (0.49) & 1.39 & 3.44 (0.28) & 0.65 & 3.59 (0.43) & 1.32 & 5.21 (0.38) \\
G327.30-0.58& 2.38 & 9.27 (0.23) & 7.72 & 33.96 (0.53) & 3.84 & 32.70 (0.38) & 1.46 & 7.56 (0.30) & 1.31 & 8.71 (0.34) & 2.30 & 15.12 (0.34) \\
G329.07-0.31& 0.82 & 3.39 (0.10) & 2.75 & 15.70 (0.10) & 2.46 & 10.88 (0.34) & - - - &- - - & 0.47 & 1.52 (0.23) & 0.77 & 2.83 (0.25)  \\
G331.51-0.34& 0.80 & 0.64 (0.14) & 1.64 & 03.86 (0.19) & 2.93 & 07.51 (0.22) & - - - &- - - & - - - &- - - &- - - &- - -\\
G331.71+0.58& 1.02 & 3.26 (0.25) & 2.90 & 15.54 (0.38) & 2.43 & 16.12 (0.91) & 0.49 & 2.01 (0.27) & - - - &- - - \\
G331.71+0.60& 1.38 & 4.51 (0.23) & 3.80 & 15.84 (0.45) & 2.99 & 12.72 (0.49) & 0.72 & 3.28 (0.28) & 0.64 & 2.06 (0.27) & 0.95 & 4.04 (0.30)\\
G332.35-0.44& 0.84 & 1.42 (0.19) & 2.15 & 05.59 (0.24) & 3.66 & 11.01 (0.26) & - - - &- - - & - - - &- - - &- - - &- - - \\
G332.47-0.52& 1.11 & 2.51 (0.20) & 3.91 & 14.61 (0.26) & 3.51 & 21.62 (0.98) & 1.16 & 3.48 (0.27) & 0.76 & 2.64 (0.26) & 1.37 & 5.36 (0.29) \\
G332.60-0.17& 0.64 & 1.41 (0.24) & 2.60 & 08.88 (0.28) & 3.17 & 11.83 (0.72) &  - - - &- - - & - - - &- - - &- - - &- - - \\
G333.13-0.56& 1.20 & 4.97 (0.11) & 3.33 & 12.23 (0.11) & 3.45 & 16.55 (0.57) &  - - - &- - - & - - - &- - - &- - - &- - -\\
G333.32+0.10& 1.41 & 3.57 (0.14) & 4.54 & 17.26 (0.14) & 3.90 & 19.45 (1.22) & - - - &- - - & 0.56 & 1.12 (0.25) & 0.83 & 2.76 (0.33)\\
G338.92+0.55& 2.05 & 9.45 (0.39) & 5.93 & 51.66 (0.77) & 6.37 & 31.78 (0.50) & 1.07 & 7.87 (0.44) & 0.63 & 4.78 (0.46) & 0.98 & 7.72 (0.47) \\
G340.97-1.02& 1.53 & 5.64 (0.26) & 5.45 & 24.75 (0.36) & 7.60 & 38.63 (0.46) & 1.45 & 6.53 (0.33) & 0.69 & 2.64 (0.27) & 1.59 & 6.93 (0.31)\\
G343.50+0.03& 1.03 & 2.39 (0.19) & 3.75 & 11.41 (0.22) & 3.63 & 13.03 (0.23) & 0.65 & 1.88 (0.24) & - - - &- - - &- - - &- - - \\

  \hline
  HII regions \\
  \hline
G12.42+0.50& 0.66 & 1.98 (0.20) & 2.99 & 10.27 (0.22) & 8.33 & 24.25 (0.46) & 1.77 & 4.85 (0.22) & 1.01 & 3.35 (0.25) & 1.67 & 5.50 (0.23)\\
G310.15+0.76& 0.53 & 1.22 (0.22) & 1.78 & 06.53 (0.27) & 2.49 & 20.99 (0.43) & 1.27 & 2.97 (0.23) & 0.67 & 2.16 (0.26) & 1.38 & 4.18 (0.25) \\
G320.23-0.28& 1.12 & 3.10 (0.24) & 3.12 & 12.77 (0.34) & 2.58 & 17.46 (0.40) & 0.82 & 3.11 (0.31) & 0.89 & 3.43 (0.29) & 1.44 & 6.04 (0.31)\\
G327.40+0.44& 0.91 & 4.78 (0.40) & 2.78 & 12.01 (0.68) & 2.51 & 15.52 (0.64) & 0.58 & 3.05 (0.35) & 0.52 & 2.02 (0.30) & 0.87 & 3.51 (0.32) \\
G330.88-0.37& 0.34 & 2.38 (0.39) & 1.73 & 07.67 (0.40) & 5.62 & 37.74 (0.39) & 1.02 & 4.81 (0.33) & 0.96 & 5.47 (0.34) & 1.83 & 9.76 (0.33)\\
G330.95-0.18& - - - & - - - & 1.28 & 15.14 (0.71) & 4.58 & 45.80 (3.69) & 0.89 & 7.77 (0.63) & 0.40 & 2.80 (0.50) & 1.22 & 9.87 (0.59) \\
G331.13-0.24& 0.76 & 2.98 (0.29) & 2.48 & 15.51 (0.48) & 2.29 & 20.59 (1.95) & - - - &- - - & - - - &- - - &- - - &- - - \\
G337.40-0.40& 0.71 & 3.50 (0.30) & 3.10 & 11.94 (0.24) & 3.31 & 20.29 (0.55) & 1.61 & 6.97 (0.27) & 1.03 & 4.58 (0.27) & 1.99 & 8.36 (0.27) \\
G339.58-0.13& 1.04 & 3.27 (0.27) & 4.20 & 15.54 (0.30) & 1.66 & 11.47 (0.42) & - - - &- - - & - - - &- - - &- - - &- - - \\
G340.05-0.25& 1.11 & 3.68 (0.28) & 3.78 & 15.23 (0.33) & 2.90 & 23.48 (0.52) & 0.98 & 5.97 (0.39) & 0.72 & 3.89 (0.38) & 1.41 & 5.57 (0.30)\\
G345.00-0.22& 1.09 & 4.42 (0.24) & 2.94 & 12.85 (0.53) & 1.75 & 10.24 (0.37) & 0.72 & 3.94 (0.30) & 0.41 & 1.80 (0.29) & 0.72 & 4.25 (0.26) \\
 \hline
 \end{tabular}
 \label{tb:rotn}
\end{minipage}
\end{table*}

\begin{table*}
\begin{minipage}{18cm}
 \caption{\label{tab:test}Derived parameters of the molecular lines.}
 \begin{tabular}{cccccccccc}
  \hline
  \hline
 EGO  & \multicolumn{3}{|c|}{N$_2$H$^+$}&  \multicolumn{3}{|c|}{H$^{13}$CO$^+$} & \multicolumn{2}{|c|}{C$_2$H}& \\

 name   & T$_{ex}$$^a$ & $\tau$ & $N$ & T$_{ex}$ & $\tau$ & N & $\tau$ & $N$ &  \\
 \hline
 & (K) & & (10$^{13}$ cm$^{-2}$)  & (K) & & (10$^{13}$ cm$^{-2}$)  & & (10$^{14}$ cm$^{-2}$)  \\
  \hline
  MYSOs \\
  \hline
G10.34$-$0.14 & 21.2(2.1) & 0.35(0.04) & 7.5(0.9)& 7.2(1.4) & 0.23(0.03) & 2.2(0.5) & 0.88(0.21) & 4.0(0.8)\\
G14.63$-$0.58 & 14.8(0.5) & 0.56(0.08) & 6.2(0.3)& 5.5(1.1) & 0.44(0.04) & 3.8(1.0) & 1.78(0.34) & 4.8(0.5) \\
G309.38$-$0.13& 9.1(0.4) & 0.56(0.07) & 3.0(0.2)& - - - & - - - & - - - & - - - & - - - \\
G309.91+0.32& 15.4(1.0) & 0.49(0.06) & 4.6(0.4)& - - - & - - - & - - - & - - - & - - - \\
G313.76-0.86& 15.3(7.9) (A) & 0.49(0.04) & 7.2(3.8)& 9.9(2.0) & 0.21(0.03) & 5.1(1.5) & 0.36(0.10) & 5.1(1.9) \\
G326.41+0.93& 15.3(7.9) (A) & 0.57(0.10) & 6.7(3.4)& - - - & - - - & - - - & - - - & - - - \\
G326.48+0.70& 11.5(0.2) & 1.30(0.17) & 11.2(0.4)& 9.3(1.8) & 0.25(0.04) & 3.8(1.0) & 0.88(0.21) & 5.0(0.5) \\
G327.30-0.58& 16.6(1.4) & 0.85(0.08) & 15.8(1.6)& 6.9(1.4) & 0.47(0.06) & 9.6(2.1) & 0.54(0.18) & 15.1(3.5) \\
G329.07-0.31& 18.1(8.1) & 0.20(0.03) & 5.7(2.5)& - - - & - - - & - - - & 0.90(0.27) & 3.7(1.0) \\
G331.51-0.34& 15.3(7.9) (A) & 0.14(0.08) & 4.9(2.3)& - - - & - - - & - - - & - - - & - - - \\
G331.71+0.58& 7.4(0.5) & 1.10(0.12) & 6.6(0.6)& 6.0(1.2) & 0.18(0.02) & 2.3(0.8) & - - - & - - - \\
G331.71+0.60& 9.2(0.3) & 0.98(0.05) & 6.3(0.4)& 6.0(1.2) & 0.27(0.03) & 3.9(1.1) & 1.45(0.43) & 4.4(0.5) \\
G332.35-0.44& 7.8(0.8) & 0.59(0.05) & 1.9(0.4)& - - - & - - - & - - - & - - - & - - - \\
G332.47-0.52& 18.2(7.5) & 0.30(0.17) & 5.6(2.3)& 7.1(1.4) & 0.37(0.04) & 4.0(1.1) & 0.44(0.13) & 5.5(1.5) \\
G332.60-0.17& 15.3(7.9) (A) & 0.24(0.10) & 3.0(1.4)& - - - & - - - & - - - & - - - & - - - \\
G333.13-0.56& 6.7(0.4) & 2.30(0.17) & 8.2(0.6)& - - - & - - - & - - - & - - - & - - - \\
G333.32+0.10& 15.3(7.9) (A) & 0.47(0.06) & 6.5(3.1)& - - - & - - - & - - - & 1.49(0.45) & 4.4(1.8) \\
G338.92+0.55& 15.3(7.9) (A) & 0.67(0.10) & 16.4(8.2)& 9.5(1.8) & 0.18(0.03) & 8.6(2.4) & 1.14(0.34) & 9.2(2.3) \\
G340.97-1.02& 23.4(3.0) & 0.31(0.08) & 11.1(1.9)& 10.7(1.9) & 0.21(0.03) & 7.5(1.6) & 0.88(0.21) & 10.5(1.8) \\
G343.50+0.03& 33.6(3.5) & 0.13(0.05) & 6.1(0.7)& 6.6(1.3) & 0.20(0.02) & 2.1(0.6) & - - - & - - - \\

  \hline
  HII regions \\
  \hline
G12.42+0.50& 7.7(2.8) (A) & 1.10(0.15) & 4.3(1.6)& 11.4(2.2) & 0.24(0.04) & 5.6(1.7) & 0.90(0.27) & 5.1(0.4) \\
G310.15+0.76& 7.7(2.8) (A) & 0.50(0.08) & 2.1(0.8)& 6.0(1.2) & 0.55(0.06) & 4.1(1.1) & 0.48(0.19) & 2.9(0.9) \\
G320.23-0.28& 10.9(0.6) & 0.51(0.07) & 4.3(0.3)& 5.6(1.1) & 0.37(0.04) & 4.1(1.1) & 0.97(0.29) & 5.3(0.6) \\
G327.40+0.44& 6.2(0.3) & 2.20(0.13) & 8.1(0.8)& 5.8(1.1) & 0.23(0.03) & 3.6(1.0) & 0.60(0.18) & 3.0(0.4) \\
G330.88-0.37& 5.3(0.6) & 1.27(0.11) & 4.1(0.7)& 8.9(1.8) & 0.19(0.03) & 5.2(1.5) & 0.09(0.11) & 6.1(0.9) \\
G330.95-0.18& 7.7(2.8) (A) & 0.33(0.07) & 4.5(1.6)& 8.5(1.7) & 0.18(0.02) & 8.3(2.4) & 0.48(0.19) & 6.7(1.8) \\
G331.13-0.24& 7.7(2.8) (A) & 0.78(0.09) & 5.7(2.2)& - - - & - - - & - - - & - - - & - - - & \\
G337.40-0.40& 7.8(0.5) & 1.08(0.09) & 5.0(0.6)& 6.3(1.2) & 0.67(0.08) & 10.0(2.7) & 0.07(0.13) & 5.3(0.5) \\
G339.58-0.13& 31.6(7.6) & 0.16(0.04) & 8.0(1.9)& - - - & - - - & - - - & - - - & - - - & \\
G340.05-0.25& 12.5(6.1) & 0.52(0.04) & 5.3(1.7)& 6.5(1.3) & 0.33(0.04) & 7.2(2.1) & 0.11(0.21) & 4.1(1.2) \\
G345.00-0.22& 6.7(0.7) & 1.60(0.15) & 6.7(0.7)& 4.8(0.9) & 0.49(0.05) & 6.4(1.6) & 0.62(0.18) & 4.0(0.7) \\
 \hline
 \end{tabular}
 \label{tb:rotn}

  a: The $T_{ex}$ of N$_2$H$^+$ values marked with ``A'' represent the average value derived from other sources.
\end{minipage}
\end{table*}

\section{Introduction}
Massive stars play an important role in the evolution of galaxies.
They deposit large amounts of radiation and inject kinetic energy
into their surrounding interstellar medium (ISM) in the form of
outflows and stellar winds during their short lives. At the end of
their lives they release various heavy elements into the cosmic
space by means of supernova explosions. Understanding the formation
of massive star is thus important to our knowledge of galaxy
evolution. However, their formations are far less clear than their
low-mass counterparts. Because they are rare and most of them have
large distances ($>$ 1 kc) from our solar system. Besides, young
massive stars tend to form in clusters or groups with high dust
extinction, which further makes observations of them challenging. In
the last two decades, a lot of research has been done in
astrophysics to understand the formation of massive stars and the
feedback to their surrounding ISM (e.g. Zinnecker et al. 2007;
Deharveng et al. 2010, and references therein). It is generally
accepted that massive stars evolve through starless cores in
infrared dark clouds (IRDCs) to hot cores with central young stellar
objects, then to hypercompact and ultracompact HII regions, the
final stages are compact and classical HII regions. However, these
divisions of the process of massive star formation are rough. Many
questions are still unresolved. More observations are needed for
detailed understanding of their formation, and it is also essential
to identify their chemical properties in different evolutionary
stages.

To date many attempts have been done to search for massive star
formation regions, using data from IRAS (e.g. Molinari et al.
1996), MSX (e.g. Lumsden et al. 2002; 2013) and/or Spitzer (e.g.
Cyganowski et al. 2008) sky surveys. Cyganowski et al. (2008)
identified more than 300 Galactic extended 4.5 $\mu$m sources,
naming extended green objects (EGOs) or ``green fuzzies'' for the
common color coding of the 4.5 $\mu$m band is green in Spitzer
Infrared Array Camera (IRAC) three-color images. Although the
exact nature of this enhanced emission is still uncertain (e.g.
Reach et al. 2006; De Buizer $\&$ Vacca 2010), EGOs seem to be
mostly related to massive YSOs as the majority of them are
associated with infrared dark clouds (IRDCs) and/or Red MSX
Sources (RMSs) (Lumsden et al. 2002). Molecular line observations
support that EGOs indeed trace a population with ongoing outflow
activity and the central young protostars are still embedded in
infalling envelopes (e.g. Chen et al. 2010; Cyganowski et al.
2011; He et al. 2012; Yu $\&$ Wang 2013 ). In our sample of EGO
clumps, 26 ($\sim$ 84$\%$) are found to be associated with IRDCs.
IRDCs provide us with the possibility to investigate the
subsequently early stages of high mass star formation. Molecular
line and dust continuum observations of IRDCs have shown that they
are cold ($<$ 30 k), dense (n(H$_2$) $>$ 10$^5$ cm$^{-3}$,
N(H$_2$) $>$ 10$^{22}$ cm$^{-3}$), and massive (in the range of
10$^2$ and 10$^5$ M$_\odot$) structures with sizes of several
parsec (e.g. Carey et al. 2000). The $Spitzer$ GLIMPSE and MIPSGAL
surveys indicate different evolutionary stages of IRDCs (e.g.
Churchwell et al. 2009; Carey et al. 2009). Chambers et al. (2009)
proposed an evolutionary sequence in which ``quiescent'' clumps
(contain no IR-Spitzer emission) evolve into ``intermediate''
(EGOs without 24 $\mu$m emission), ``active'' (EGOs with 24 $\mu$m
emission) and ``red'' clumps (with bright 8$\mu$m emission,
probably HII regions).

Chemical composition of molecular gas is also thought to evolve
due to the physical changes that occur during star formation
processes. As material collapses and gets ionized by central young
stars, densities and temperature rise, leading to the production
and destruction of different molecular species. Although several
studies have been carried out to investigate whether the
evolutionary stages defined by Chambers et al. (2009) chemically
distinguishable (e.g. Vasyunina et al. 2011; Sanhueza et al. 2012;
Hoq et al. 2013; Miettinen 2014), chemical properties of massive
star formation regions are still far less explored than their
low-mass counterparts. Some observations (e.g. Hoq et al. 2013)
indicate chemical processes in massive star formation may differ
from those of low-mass star formation, as their physical
conditions (such as density, temperature) are very different. With
the aim to better understand the chemical evolution of massive
star formation, we made a molecular line study toward 31 EGO
clumps in the southern sky. We introduce our data and source
selections in Section 2, results and analysis are given in Section
3 and 4, finally we summarize in Section 5.

\begin{figure*}
\centerline{\psfig{file=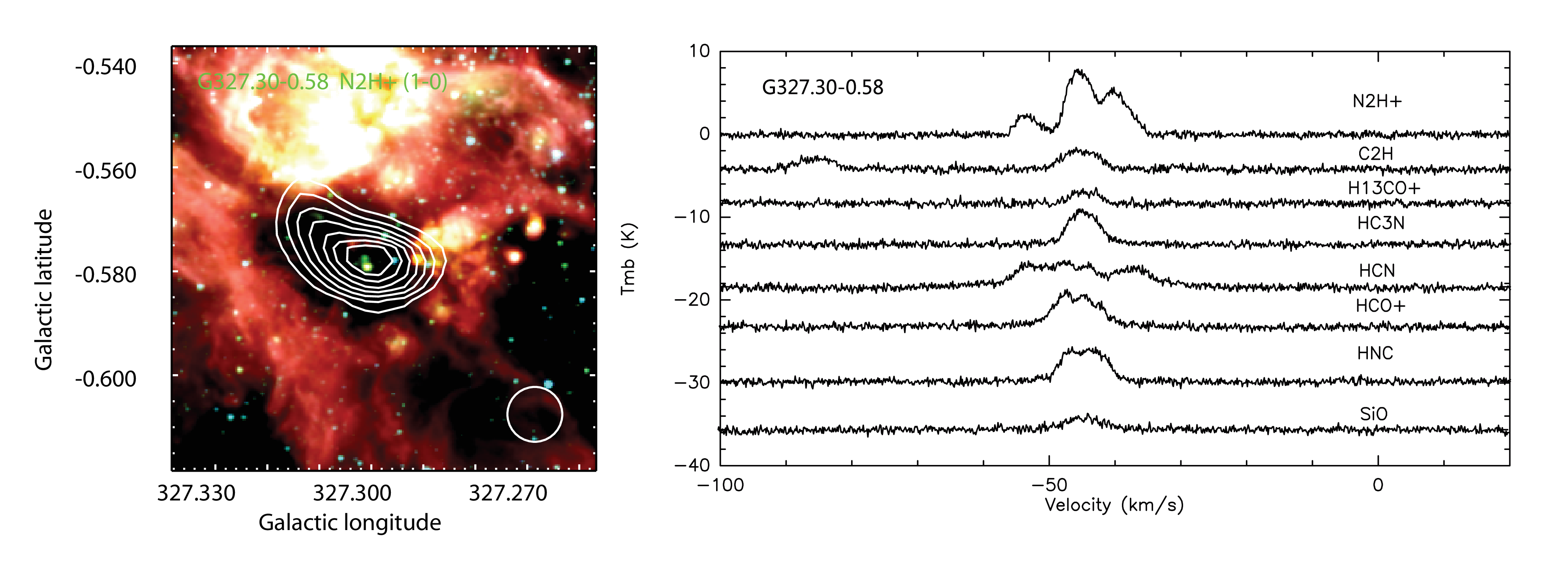,width=6.5in,height=2.5in}}
\caption{An example of the Spitzer image and the detected lines of
our sources. The white contours are N$_2$H$^+$ integrated intensity
image from MALT90. Contour levels are 30, 40,. . . ,90 percent of
the peak emission. The white circle shows the telescope beam size.
The Spitzer images of other sources can be found in the ATLASGAL
Database Server.}
\end{figure*}

\begin{figure*}
\centerline{\psfig{file=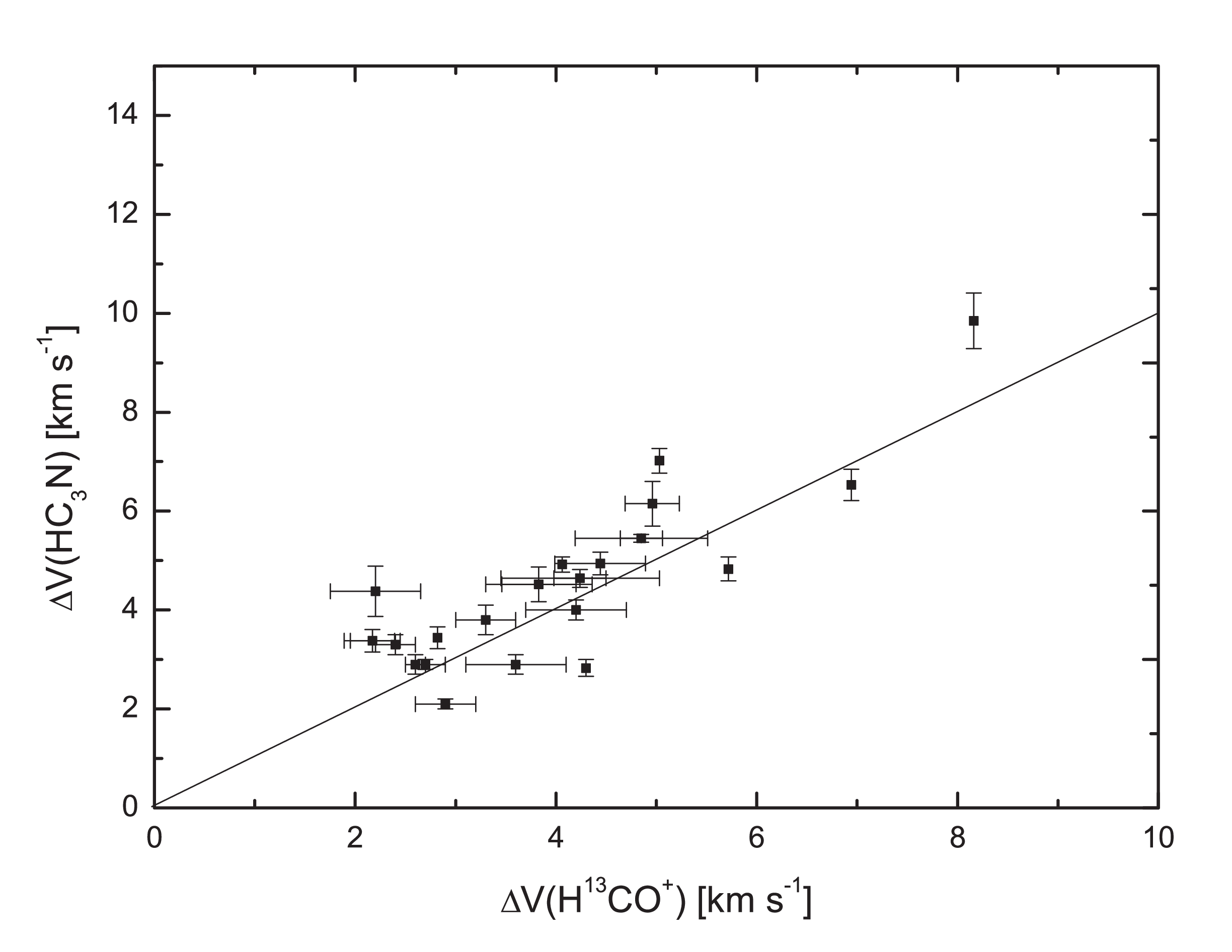,width=2.6in,height=2.0in}
\psfig{file=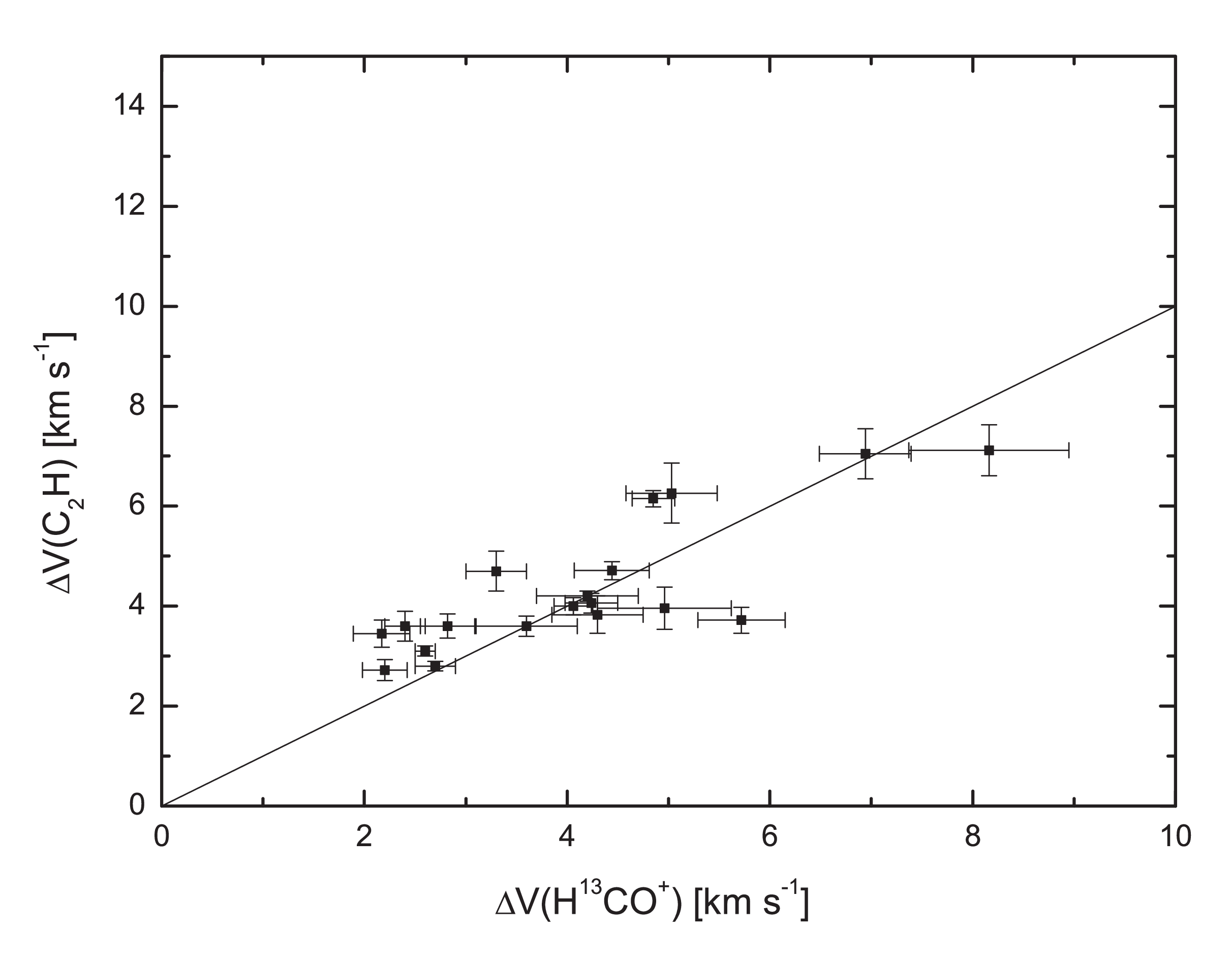,width=2.6in,height=2.0in}
\psfig{file=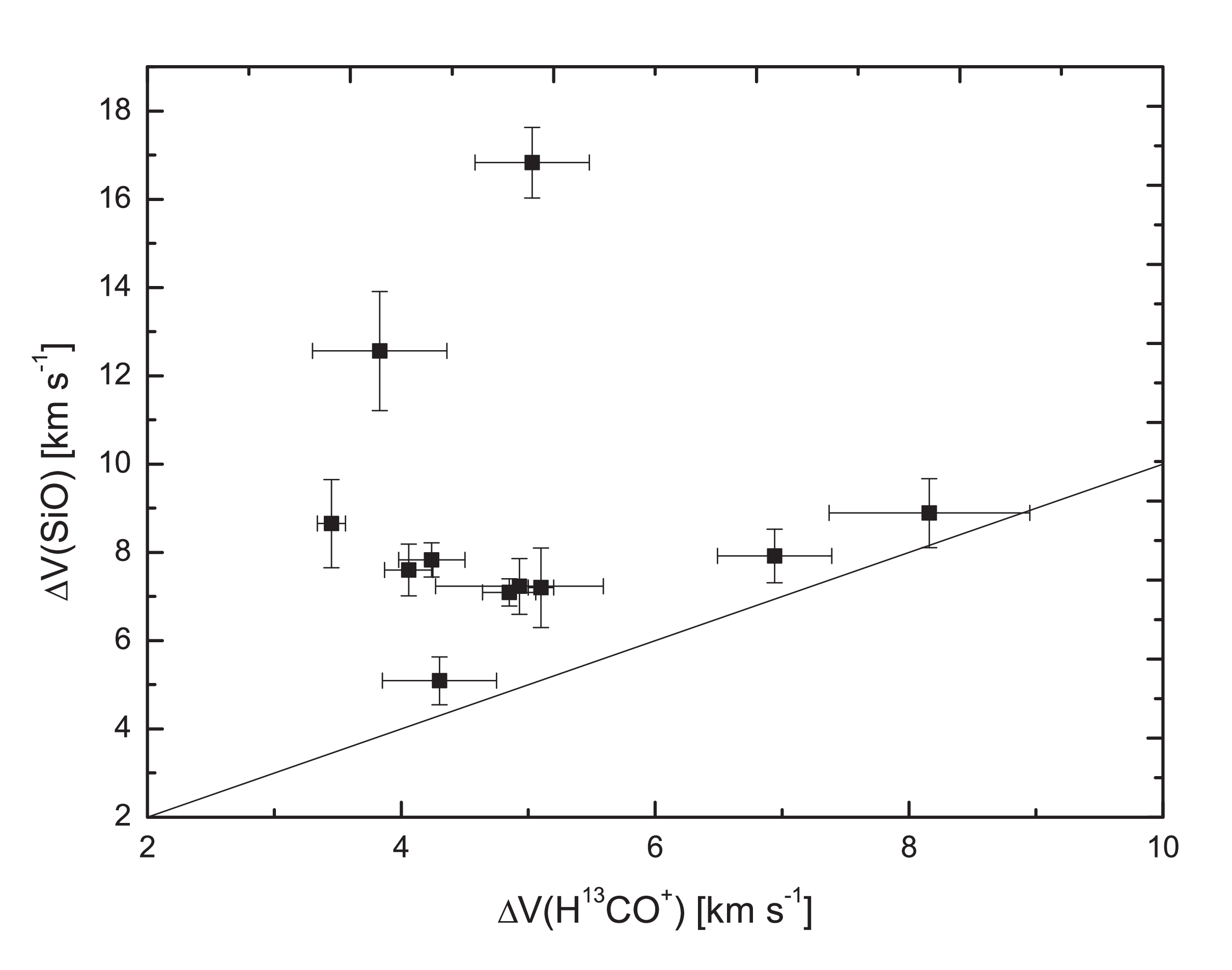,width=2.6in,height=2.0in}}
\centerline{\psfig{file=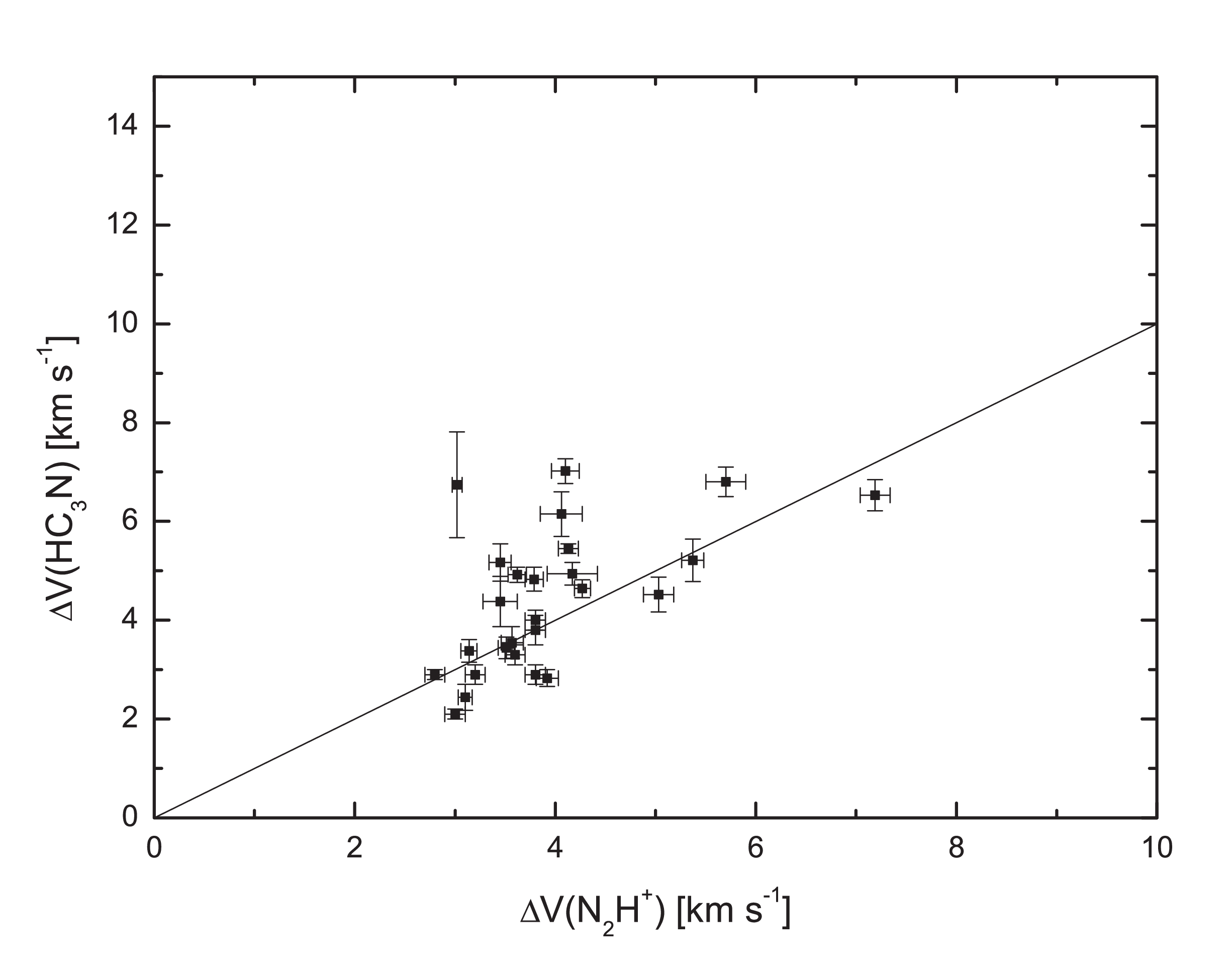,width=2.6in,height=2.0in}
\psfig{file=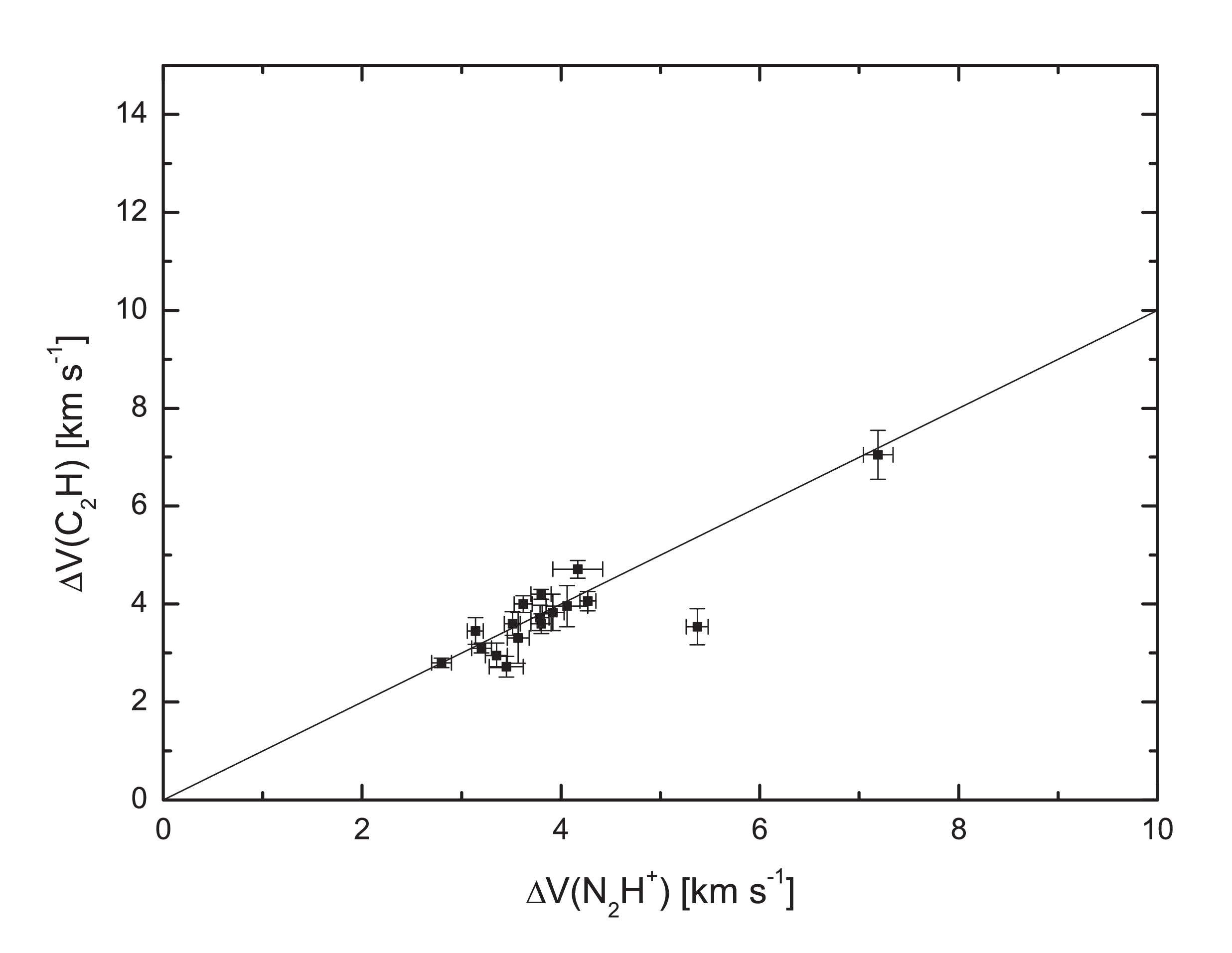,width=2.6in,height=2.0in}
\psfig{file=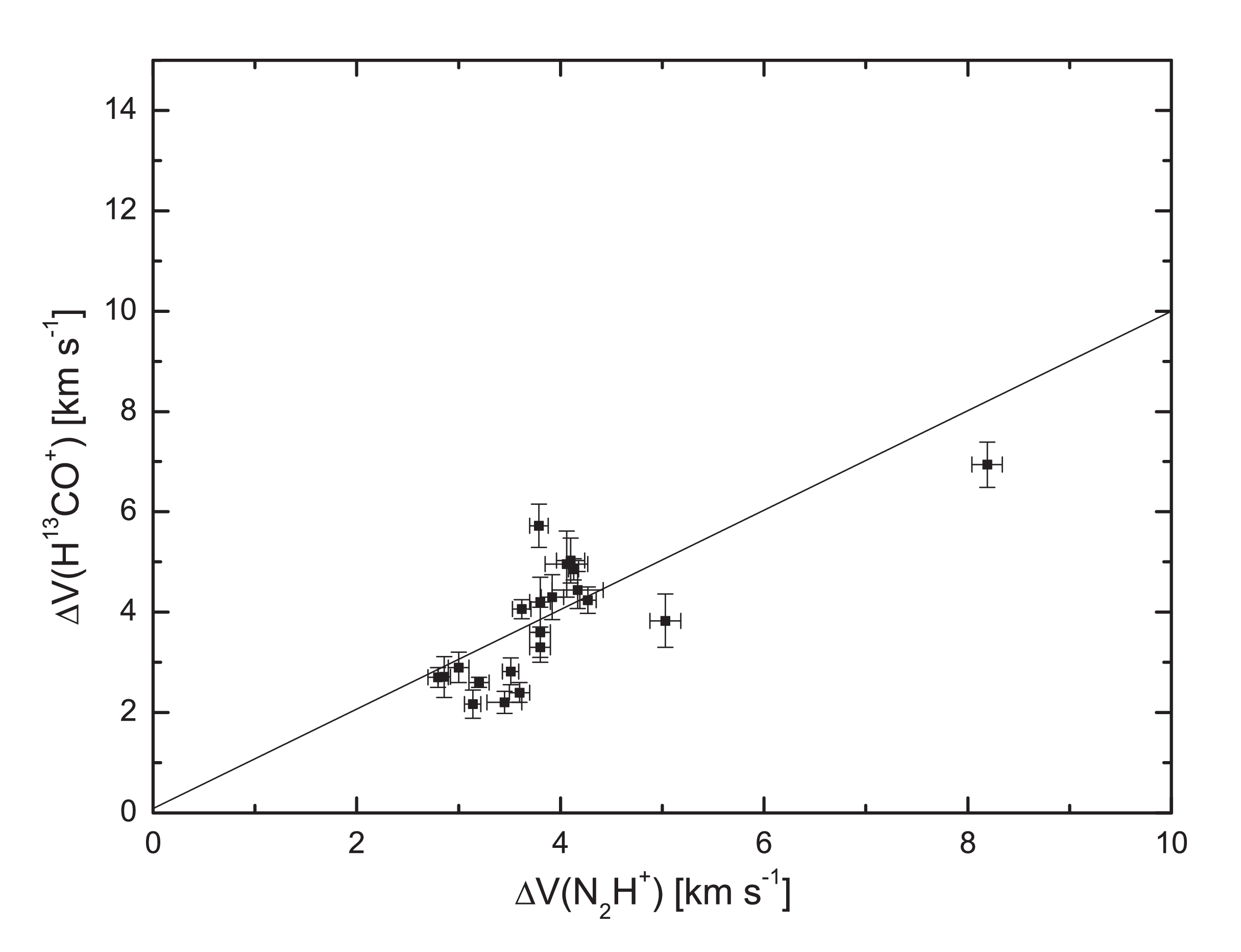,width=2.6in,height=2.0in}} \caption{ Plots
of the velocity widths of several molecular lines against those of
H$^{13}$CO$^+$ (top panels) and N$_2$H$^+$ (bottom panels). The
black lines indicate unity.}
\end{figure*}

\begin{figure*}
\centerline{\psfig{file=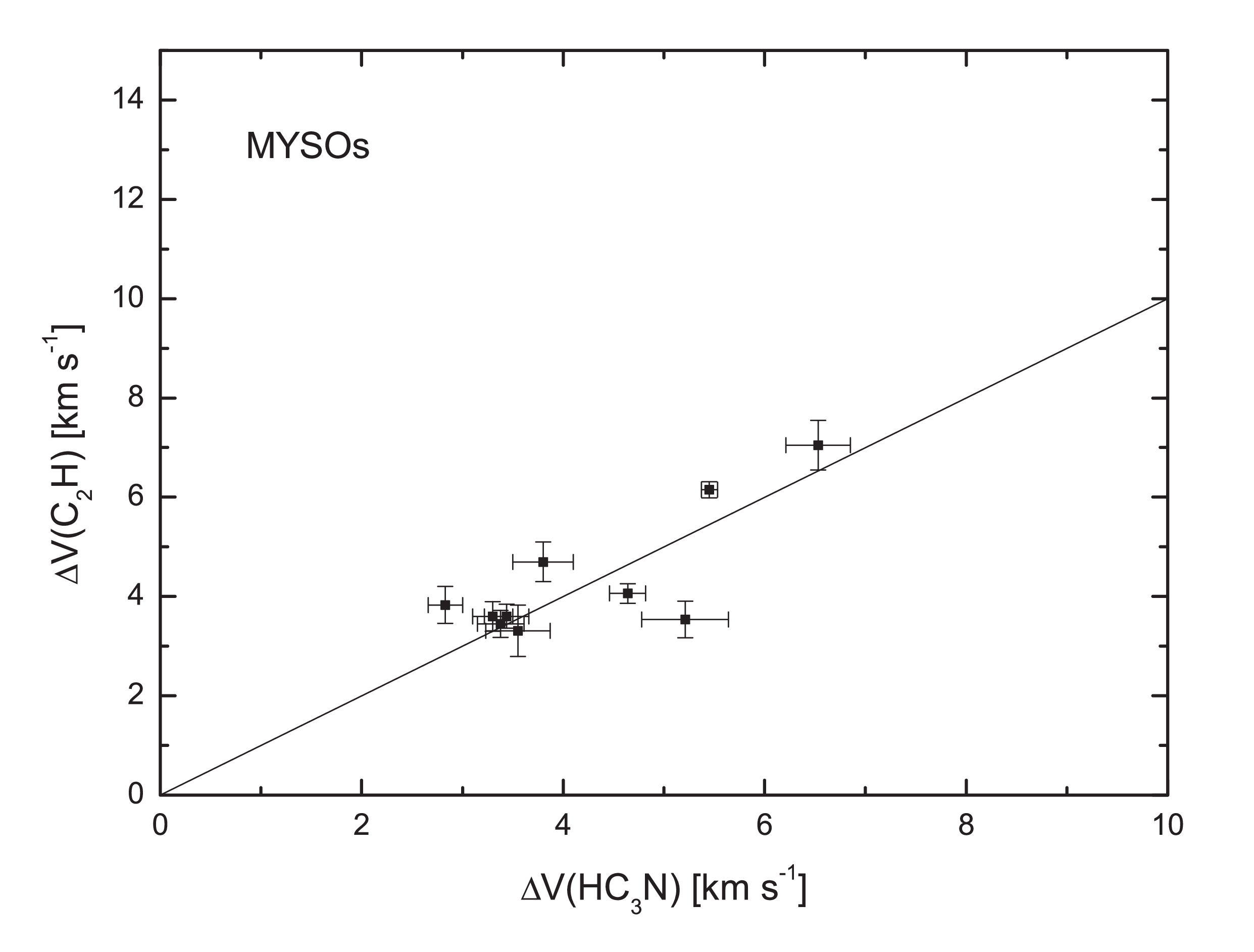,width=2.6in,height=2.0in}
\psfig{file=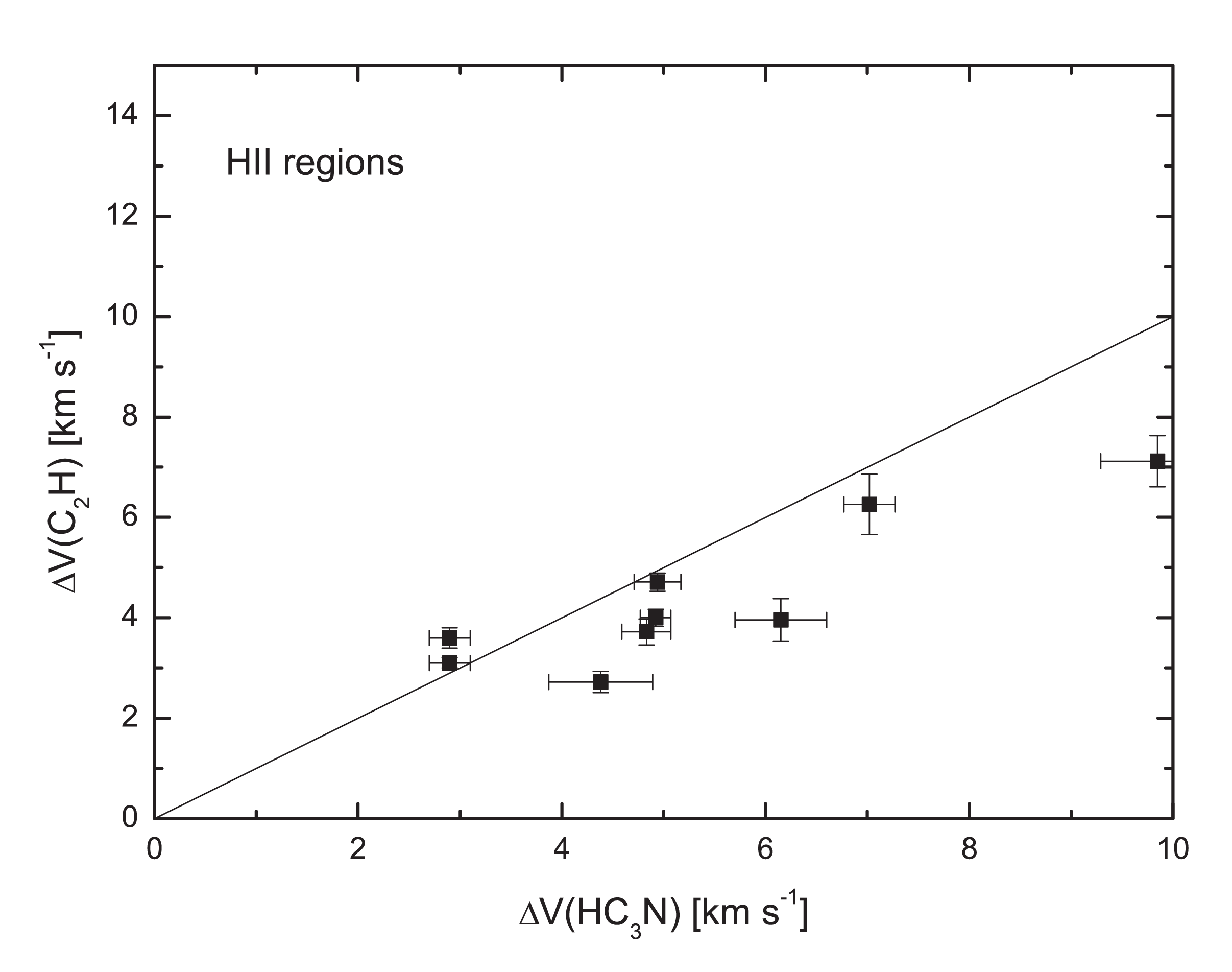,width=2.6in,height=2.0in}}
\centerline{\psfig{file=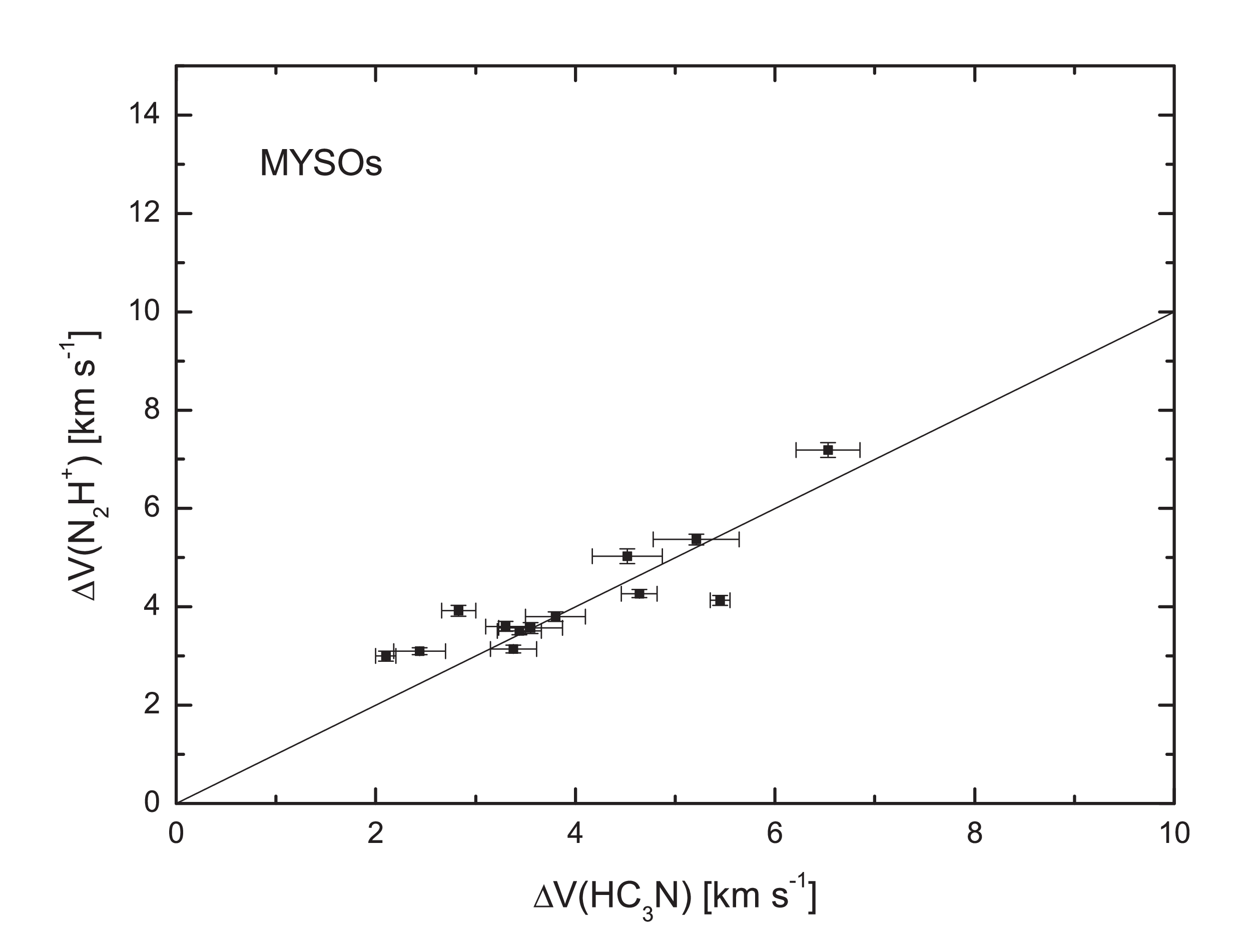,width=2.6in,height=2.0in}
\psfig{file=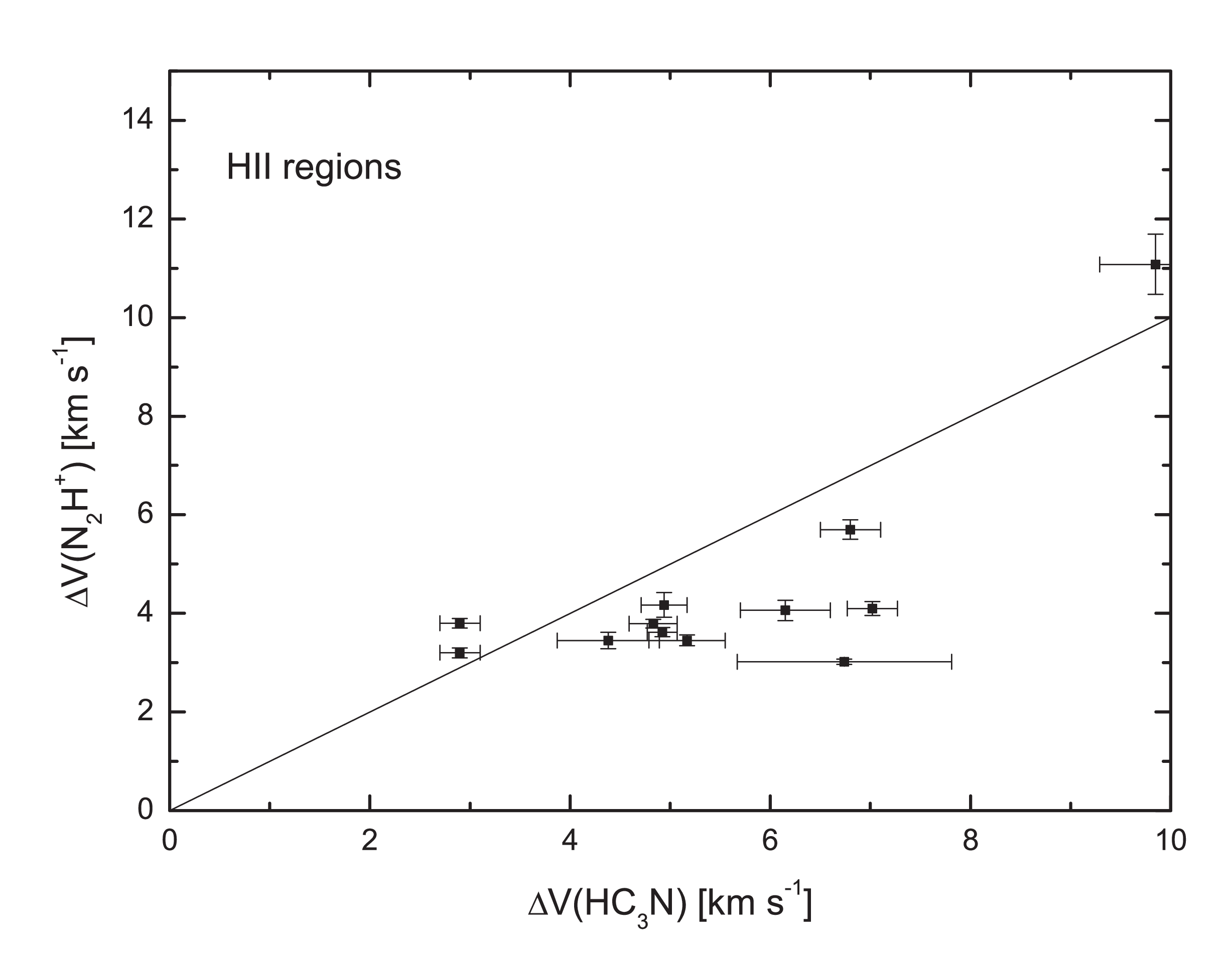,width=2.6in,height=2.0in}} \caption{Top:
Velocity width of C$_2$H against that of HC$_3$N in MYSOs and HII
regions. Bottom: Velocity width of N$_2$H$^+$ against that of
HC$_3$N in MYSOs and HII regions. The black lines indicate unity. }
\end{figure*}

\begin{figure*}
\centerline{\psfig{file=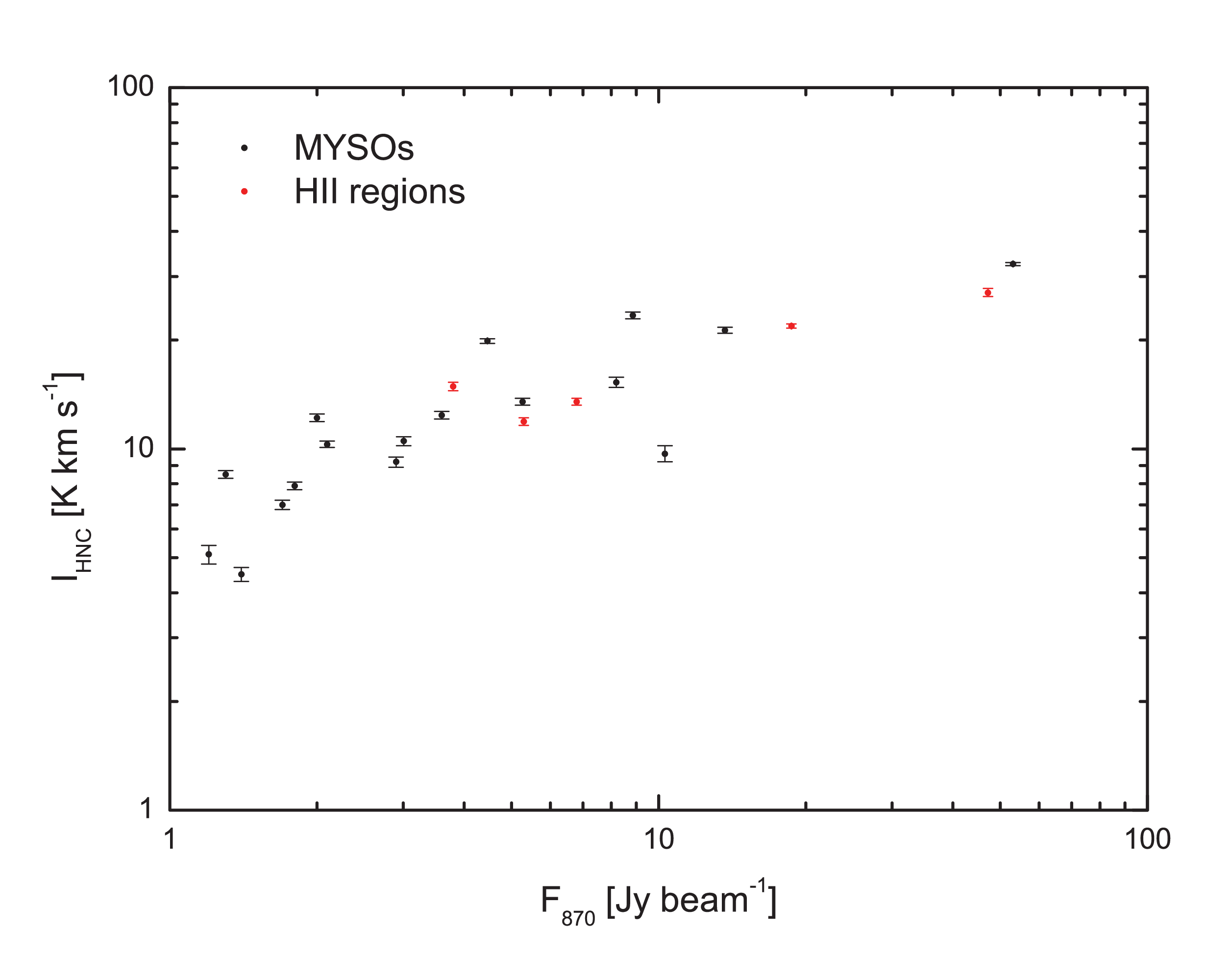,width=2.6in,height=2.0in}
\psfig{file=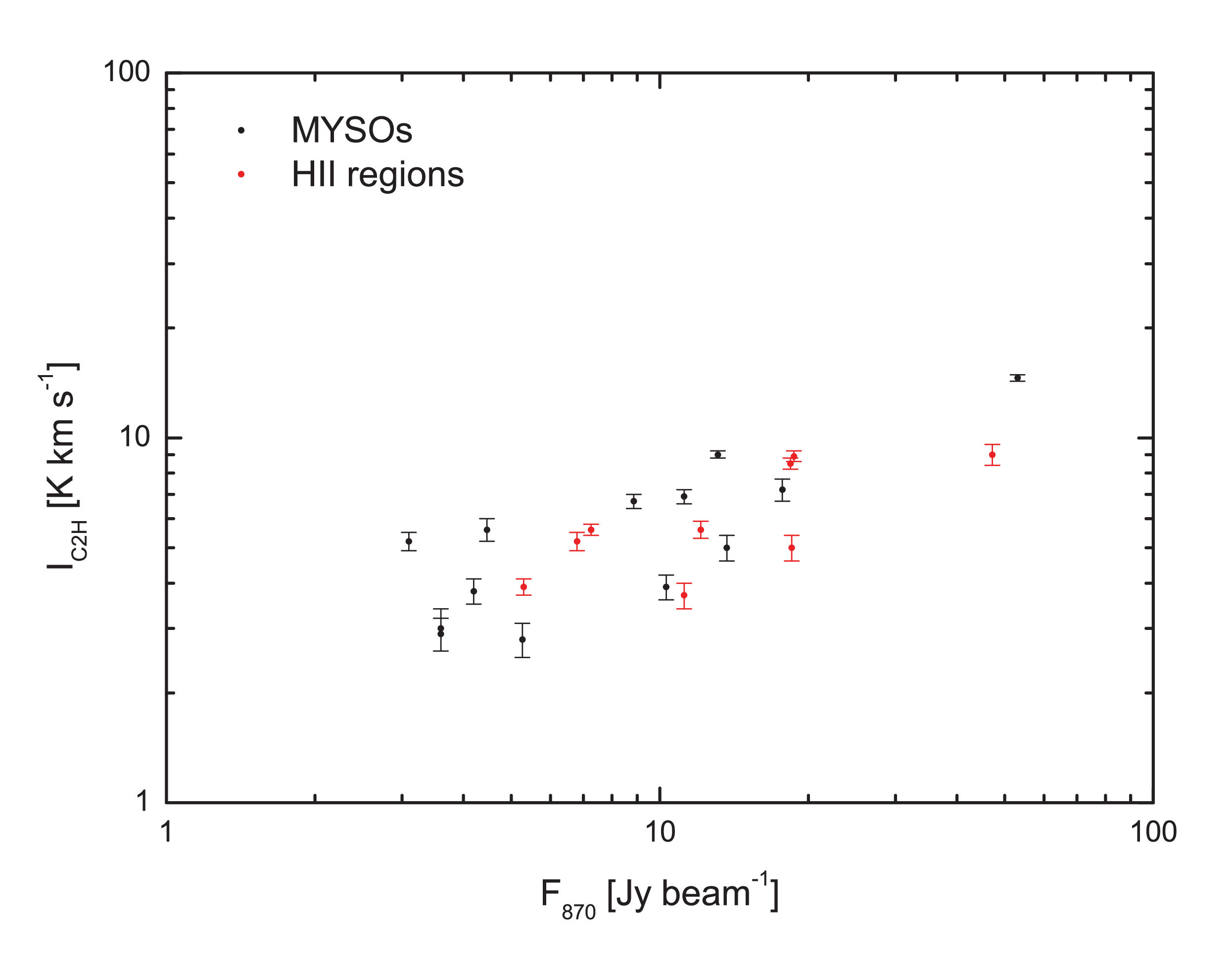,width=2.6in,height=2.0in}}
\centerline{\psfig{file=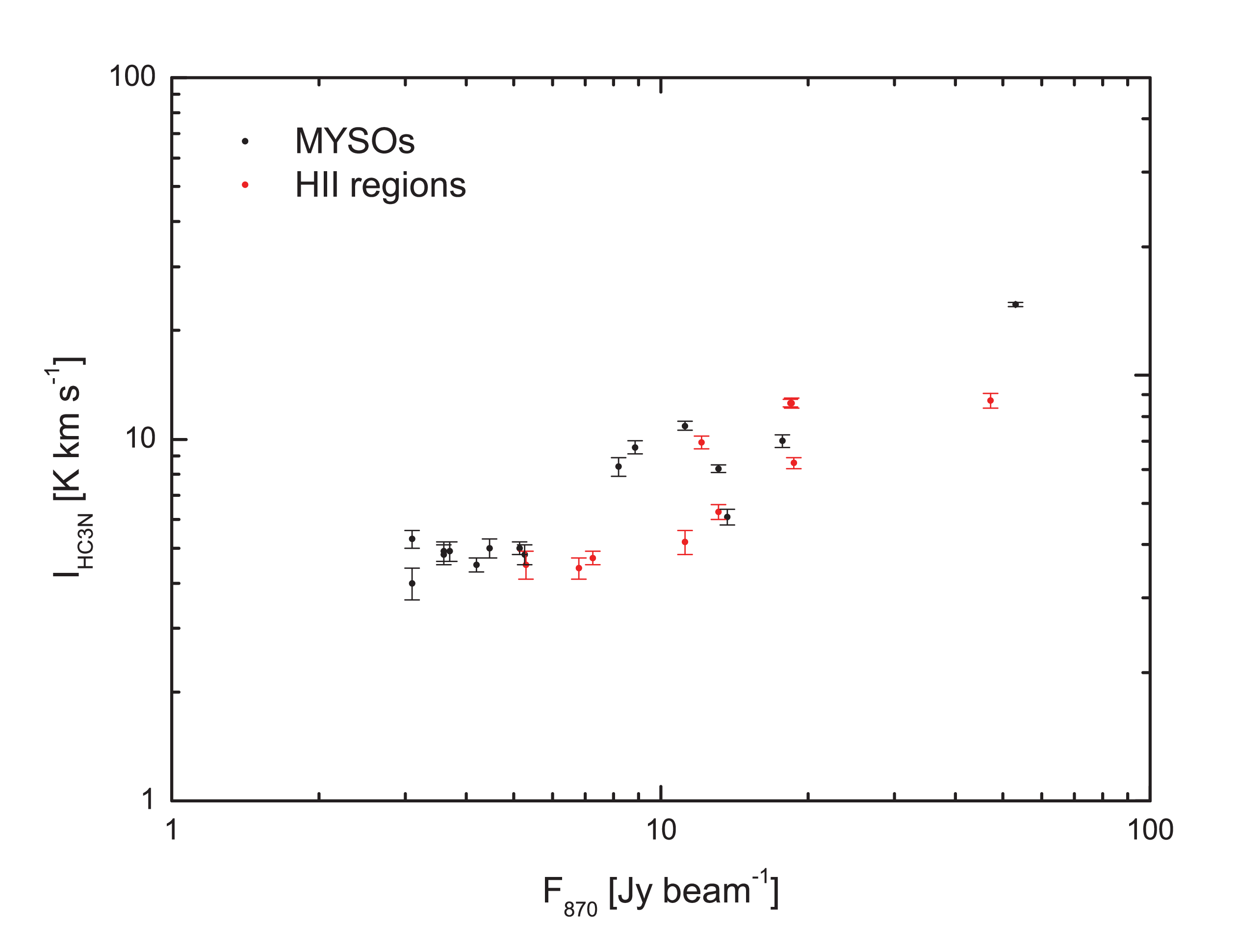,width=2.6in,height=2.0in}
\psfig{file=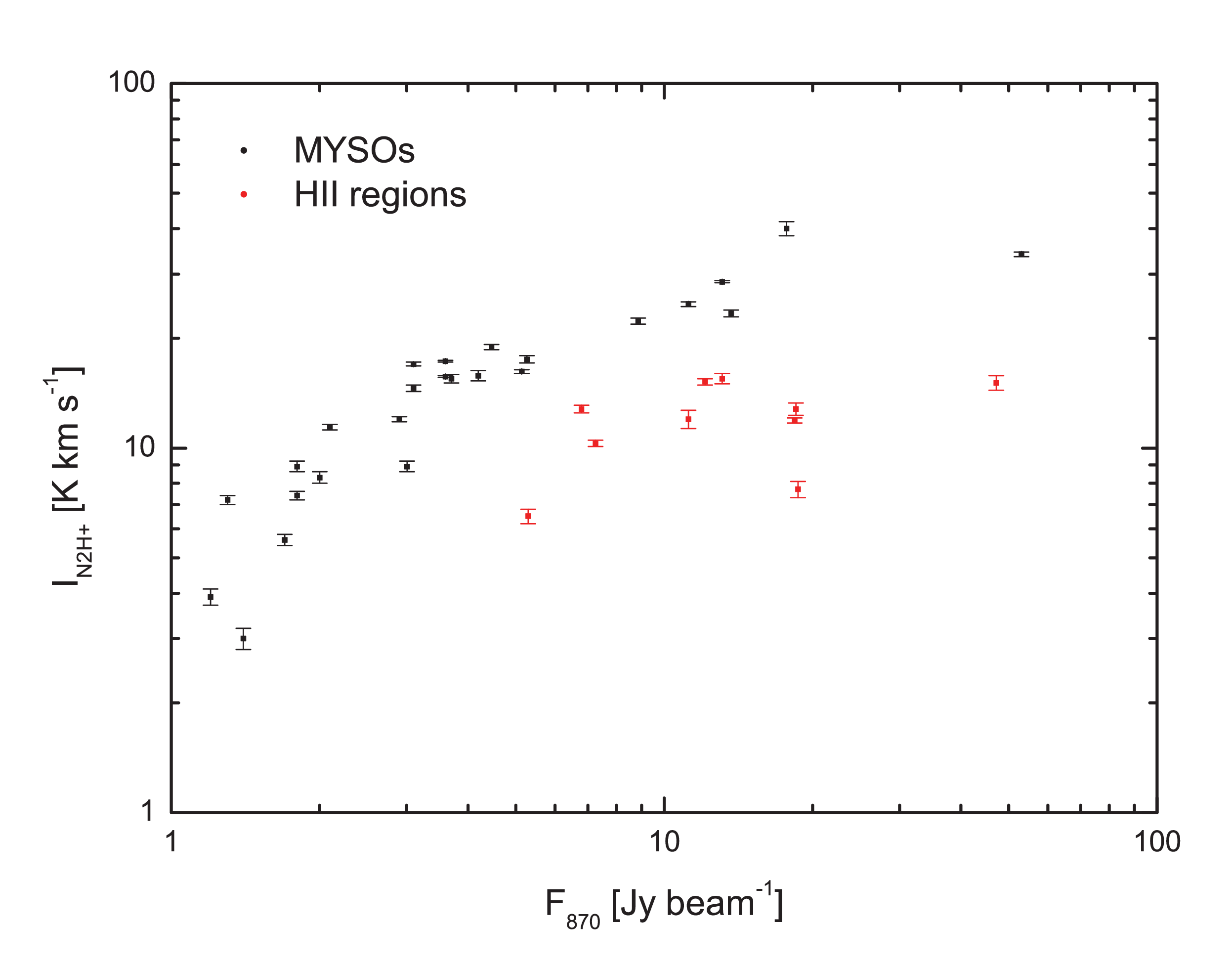,width=2.6in,height=2.0in}} \caption{The
integrated intensities of HNC, C$_2$H, HC$_3$N and N$_2$H$^+$
against the 870 $\mu$m peak flux from Contreras et al. (2013) and
Urquhart et al. (2014). }
\end{figure*}

\section{data and source selections}
Our multi-molecular line data is from MALT90. The MALT90 is a large
international project aimed at characterizing the sites within our
Galaxy where high-mass stars will form. The target clumps are
selected from the 870 $\mu$m ATLASGAL (APEX Telescope Large Area
Survey of the Galaxy) to host the early stages of high-mass star
formation. Exploiting the unique broad frequency range and
fast-mapping capabilities of the Mopra 22-m telescope, MALT90 maps
16 emission lines simultaneously at frequencies near 90 GHz. These
molecular lines will probe the physical, chemical properties and
evolutionary states of dense high-mass star-forming clumps. The
survey covers a Galactic longitude range  of $\sim$ -60 to $\sim$
15$^{\circ}$ and Galactic latitude range of -1 to +1 $^{\circ}$. The
observations were carried out with the newly upgraded Mopra
Spectrometer (MOPS). The full 8 GHz bandwidth of MOPS was split into
16 zoom bands of 138 MHz each providing a velocity resolution of
$\sim$ 0.11 km s$^{-1}$. The angular resolution of Mopra is about 38
arcsec, with beam efficiency between 0.49 at 86 GHz and 0.42 at 115
GHz (Ladd et al. 2005). The maps were made with 9$^{\prime\prime}$
spacing between adjacent rows. The MALT90 data includes (\emph{l},
\emph{b}, \emph{v}) data cubes and (\emph{l}, \emph{b}) moment and
error maps, and is publicly available from the MALT90 Home
Page\footnote{See
  http://atoa.atnf.csiro.au/MALT90}. More information
about this survey can be found in Foster et al. (2011) and Jackson
et al. (2013). The data processing was conducted using CLASS
(Continuum and Line Analysis Single-Disk Software) and GreG
(Grenoble Graphic) software packages.

With the aim to better understand the chemical evolution of
massive star formation, we made a molecular line study toward 31
EGOs in the southern sky by applying the following criteria.
(i)According to the work of Contreras et al. (2013) and Urquhart
et al. (2014), the effective diameter of an EGO clump should be
larger than the Mopra beam size ($\sim$ 38 arcsec). (ii) At least
one molecular line emission should be detected by MALT90. (iii)If
an EGO clump is associated with a RMS source, it should be
classified as ``YSO" or ``HII" by Urquhart et al. (2007).
(iv)Sources should be far from bubble or known HII regions,
considering the large beam of the 22 m Mopra telescope. Our sample
involves 13 MSX dark sources and 18 RMSs. MSX dark sources are
defined by Sakai et al. (2008) as objects with 8 $\mu$m extinction
features. Generally speaking, MSX dark sources correspond to the
``intermediate'' or ``active'' clumps defined by Chambers et al.
(2009). We also have checked the GLIMPSE images of these MSX dark
sources, none of them shows bright 8 $\mu$m emission, indicating
they are probably at earlier stages than HII regions. The RMS
survey was conceived at Leeds to systematically search the entire
Galaxy for MYSOs, by comparing the colors of sources from the MSX
and Two-Micron All-Sky Survey (2MASS) point sources (Lumsden et
al. 2002). Using the Australia Telescope Compact Array (ATCA),
Urquhart et al. (2007) completed the 5 GHz observations of 892 RMS
sources in the southern sky. Since radio interferometers may have
difficulty imaging extended emission in the Galactic plane, we
also have checked radio sky surveys such as NRAO VLA Sky Survey
(NVSS) (1420 MHz; Condon et al. 1998), Sydney University Molonglo
Sky Survey (SUMSS) (843 MHz; Mauch et al. 2003) and Westerbork
Northern Sky Survey (WENSS) (325 MHz; Rengelink et al. 1997).
\textbf{These programmes divide our RMSs into two groups:
radio-quiet and radio-loud RMSs.} Finally, we divide our sample
into two groups: MYSOs (13 MSX dark sources plus 7 radio-quiet
RMSs) and HII regions (11 radio-loud RMSs). The information of our
selected sources is listed in Table 1.

\section{results}
The most detected lines are N$_2$H$^+$ ($J=1-0$), HCO$^+$ ($J=1-0$),
HNC ($J=1-0$), HCN ($J=1-0$), HC$_3$N ($J=10-9$), H$^{13}$CO$^+$
($J=1-0$), C$_2$H ($N=1-0$) and SiO ($J=2-1$) in our sources. Their
detection rates are 100$\%$, 100$\%$, 100$\%$, 100$\%$, 71$\%$,
58$\%$, 61$\%$ and 33$\%$, respectively. N$_2$H$^+$ is a good tracer
of dense gas in the early stages of star formation as it is more
resistant to freeze-out on grains than the carbon-bearing species
(Bergin et al. 2001). Although N$_2$H$^+$ ($J=1-0$) has 15 hyperfine
transitions out of which seven have a different frequency (e.g.
Daniel et al. 2006; Pagani et al. 2009), only three groups can be
resolved in high-mass star formation regions because of turbulent
line widths. In our sources all of the N$_2$H$^+$ spectra show a
three-group line profile except G330.95-0.18. To get the line center
velocity, line width and line intensity, we fitted the three groups
with three Gaussian profiles with a fixed frequency separation
between the transitions. For G330.95-0.18 we fitted a single
Gaussian profile. C$_2$H ($N=1-0$) has six hyperfine components out
of which two ($N=1-0, J=3/2-1/2, F=2-1$ and $N=1-0, J=3/2-1/2,
F=1-0$) could easily be detected in our sample (see Figure 1). Like
the method described above, we used a multi-Gaussian function to
derive its line parameters. Although C$_2$H was first detected by
Tucker et al. (1974) in interstellar clouds, it has not been
systematically studied in massive star formation regions.
Observations indicate that C$_2$H could trace photodissociation
region (PDR) (e.g. Fuente et al. 1993). Beuther et al. (2008)
proposed C$_2$H could also trace dense gas in early stages of star
formation, but later it is rapidly transformed to other molecules in
the hot-core phase. Sanhueza et al. (2012) found no clear
evolutionary trends for C$_2$H column densities and abundances in
their sample of IRDCs. On the other hand, Miettinen (2014) found
that both the mean and median values of the C$_2$H abundance are
higher towards IR-dark clumps than towards IR-bright clumps,
indicating C$_2$H tend to trace the cold gas. More observations
should be carried out to study C$_2$H in massive star formation
regions. H$^{13}$CO$^+$ and HC$_3$N are also dense gas tracers and
could always be regarded as optically thin lines as their line shape
are relatively simple in our sources. According to chemical models
the H$^{13}$CO$^+$ abundance does not vary so much with time (e.g.
Bergin et al. 1997; Nomura $\&$ Millar 2004). HC$_3$N can trace both
the cold molecular clouds and hot cores (e.g. Chapman rt al. 2009).
We have fitted a single Gaussian profile to get their line
parameters. HCO$^+$ often shows infall signatures and outflow
wings(e.g., Rawlings et al. 2004; Fuller et al. 2005). HNC is
particularly prevalent in cold gas (Hirota et al. 1998). The HCO$^+$
($J=1-0$) and HNC ($J=1-0$) lines tend to show the so-called red or
blue profiles in our sources, indicating outflow and/or large scale
infall activities. SiO is often seen when SiO is formed from shocked
dust grains, typically in outflows (Schilke et al. 1997). The
detection rate of SiO is 50$\%$ in our H$^{13}$CO$^+$ detected EGO
clumps. This value is consistent with that of He et al. (2012).
Considering the large beam of Mopra, this detection rate is just a
lower limit. HCN($J=1-0$) has three hyperfine components. In the
optically thin case, these components have relative intensities of
1:3:5. However, it is difficult to perform Gaussian fits and analyze
the HCN lines, because most HCN transitions in our sources are
blended, showing self-absorbed line profiles and extended wing
emissions.

\subsection{line widths}
Molecular line widths observed in interstellar clouds give
information about gas kinematics (such as turbulence, rotation,
outflow and infall) inside. Optical depth could also enlarge line
widths. Here we assume N$_2$H$^+$, C$_2$H, H$^{13}$CO$^+$ and
HC$_3$N are all optically thin and their line widths mainly arise
from turbulence. Figure 2 shows the line widths of several lines
against those of H$^{13}$CO$^+$ and N$_2$H$^+$. It can be noted that
the velocity widths of HC$_3$N and C$_2$H are similar to that of
H$^{13}$CO$^+$. The width of SiO is always broader than that of
H$^{13}$CO$^+$, showing a high degree of scatter. The scatter is
probably caused by molecular outflow activities. The best $\Delta$V
correlation with N$_2$H$^+$ is C$_2$H, indicating their emissions
might originate from the same region. Figure 3 shows plots of the
velocity widths of C$_2$H and N$_2$H$^+$ against those of HC$_3$N in
MYSOs and HII regions respectively. Velocity widths of C$_2$H and
N$_2$H$^+$ against H$^{13}$CO$^+$ have the similar situation and are
omitted from display here. It is apparent that the velocity widths
of these lines are comparable to each other in MYSOs. However, in
HII regions the velocity widths of N$_2$H$^+$ and C$_2$H tend to be
narrower than those of H$^{13}$CO$^+$ and HC$_3$N. Our results seem
to support that N$_2$H$^+$ and C$_2$H emissions do not come from the
stirred up gas in the center of the clumps.

\subsection{integrated intensities}
In Figure 4, we plot the integrated intensities of HNC ($J=1-0$),
C$_2$H ($N=1-0, J=3/2-1/2, F=2-1$), HC$_3$N ($J=10-9$) and
N$_2$H$^+$ (group 2 defined by Purcell et al. (2009)) against the
870 $\mu$m peak flux from Contreras et al. (2013) and Urquhart et
al. (2014). The HNC integrated intensity increases with the APEX 870
$\mu$m peak flux both in MYSOs and HII regions. The integrated
intensities of C$_2$H and HC$_3$N are not very different between
MYSOs and HII regions. However, the N$_2$H$^+$ integrated
intensities in MYSOs tends to be larger than those in HII regions.
This may be caused by the depletions of N$_2$H$^+$ in HII regions.
Both theories and observations of low mass star formation regions
indicate N$_2$H$^+$ could be destroyed by CO (N$_2$H$^+$ + CO
$\rightarrow$ HCO$^+$ + N$_2$) and/or through dissociative
recombination with electrons produced by UV photons from the central
stars (N$_2$H$^+$ + e$^-$ $\rightarrow$ N$_2$ + H or NH + N). We may
use this method to identify whether HII regions have formed in
massive star formation regions.

\begin{figure}
\psfig{file=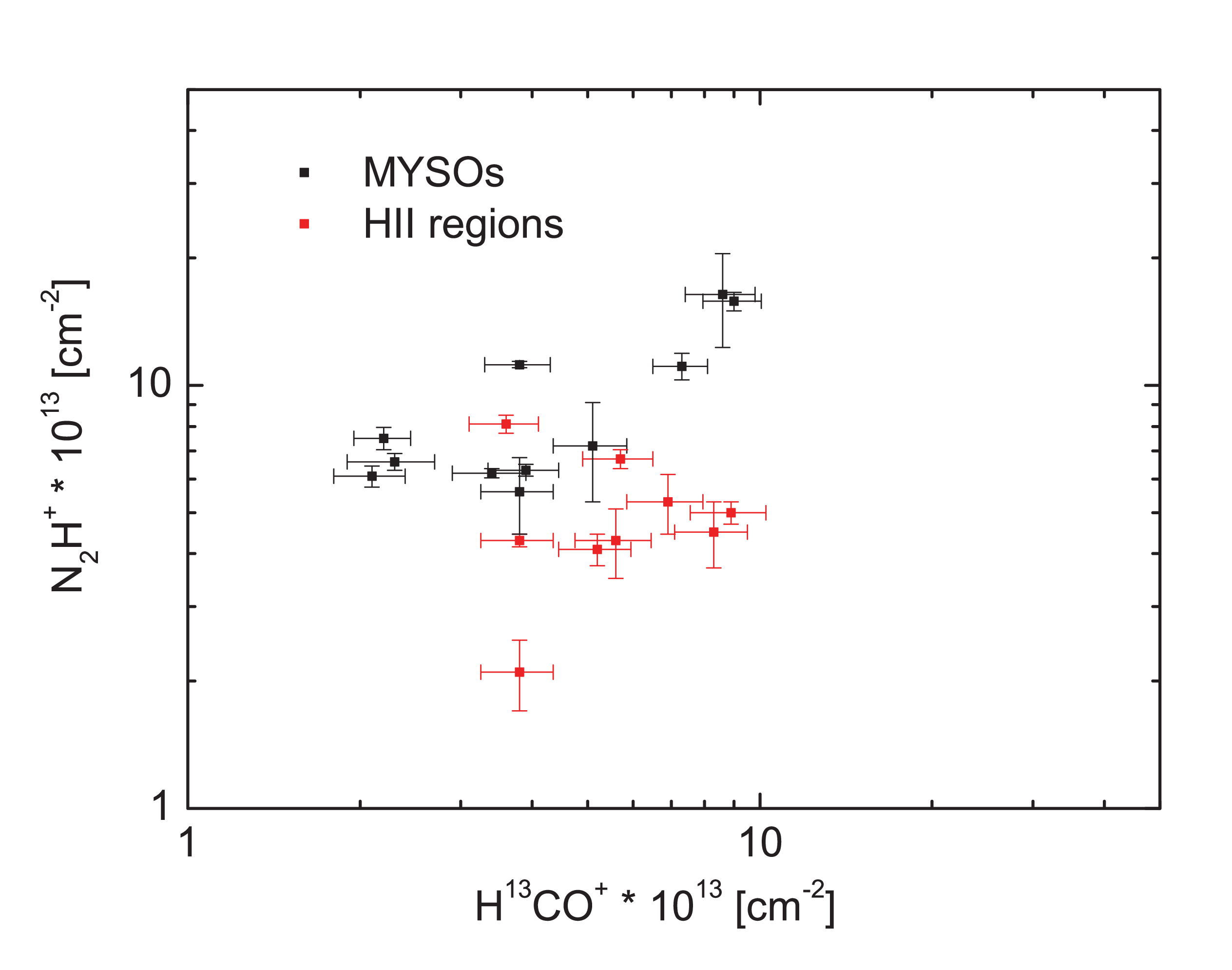,width=2.6in,height=2.0in}
\psfig{file=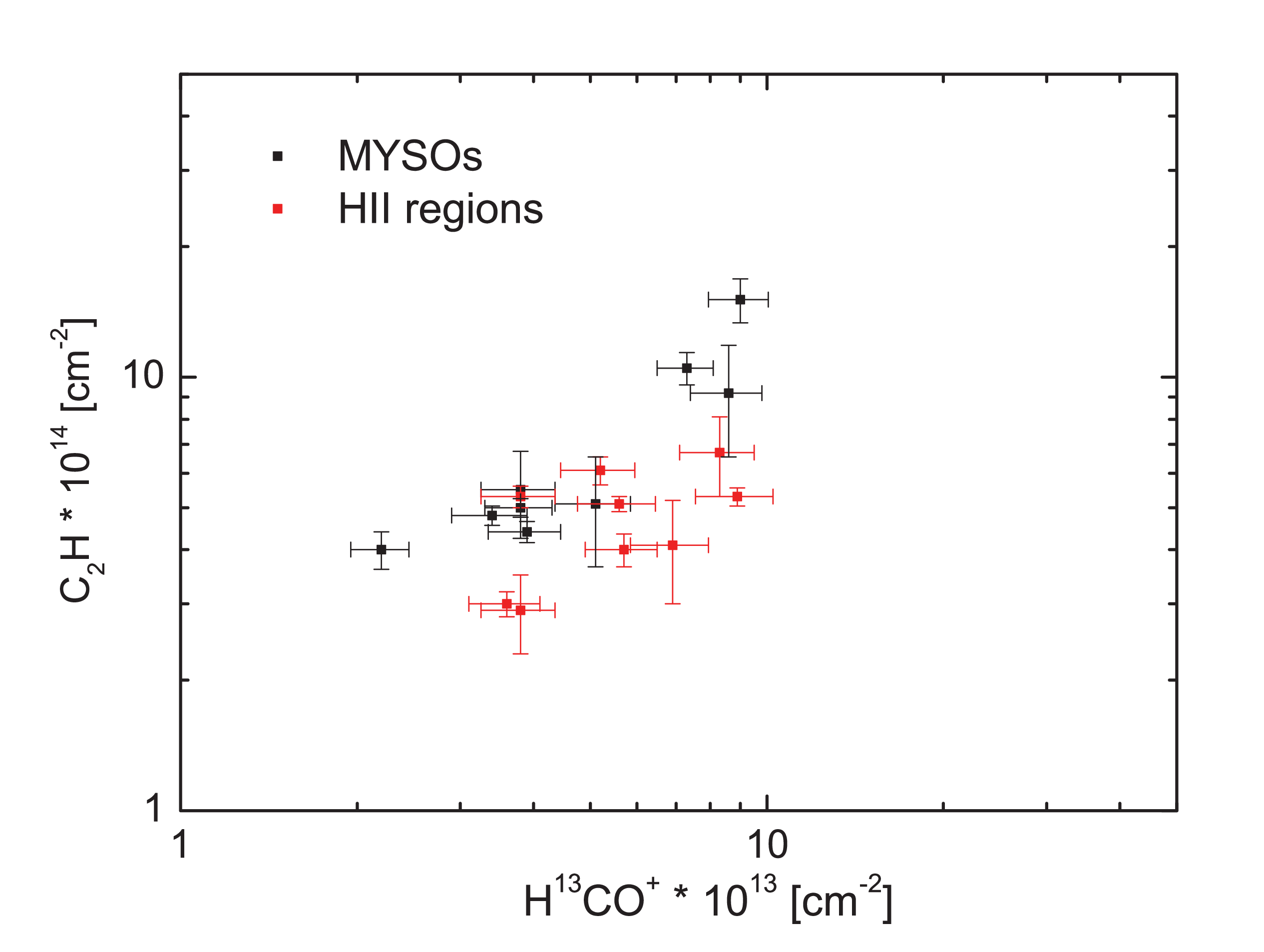,width=2.6in,height=2.0in} \caption{Top: Column
density of N$_2$H$^+$ against that of H$^{13}$CO$^+$. Bottom: Column
density of C$_2$H against that of H$^{13}$CO$^+$. }
\end{figure}

\subsection{column densities}
Even though the 4.5$\mu$m extent of an EGO is much smaller than
the Mopra beam, it is not a viable proxy for the size of the 3 mm
molecular line emission region. As the true size of our line
emission region is unknown, we assume a beam filling factor of 1
for all lines. We derived the column densities of N$_2$H$^+$,
H$^{13}$CO$^+$ and C$_2$H from the observed lines by assuming the
local thermodynamic equilibrium (LTE) conditions.

We followed the procedure outlined by Purcell et al. (2009) to
estimate the optical depth of N$_2$H$^+$. Assuming the line widths
of the individual hyperfine components are all equal, the integrated
intensities of the three blended groups should be in the ratio of
1:5:3 under optically thin conditions. The optical depth can then be
derived from the ratio of the integrated intensities of any groups
using the following equation:
\begin{equation}
\frac{\int T_{MB,1} dv}{\int T_{MB,2} dv} = \frac{1 -
exp(-\tau_1)}{1 - exp(-a\tau_2)}
\end{equation}
where $a$ is the expected ratio of $\tau_2/\tau_1$ under optically
thin conditions. We determined the optical depth only from the
intensity ratio of group 1/group 2, as anomalous excitation of the
$F_1F$ = 10-11 and 12-12 components (in our group 3) has been
reported by Caselli et al. (1995). The optical depth of group 2 is
listed in table 4 because it provides a better estimate for the
excitation temperature when $\tau
> 0.1$. Then the excitation temperature ($T_{ex}$) for N$_2$H$^+$
could be calculated with the following formula:
\begin{equation}
T_{ex} = 4.47 / ln(1+ (\frac{T_{MB}}{4.47(1 - exp(-\tau))} +
0.236)^{-1})
\end{equation}
The N$_2$H$^+$ optical depth of some clumps could not be derived
through this method. Because their ($\int$$T_{MB,1}$
$dv$)/($\int$$T_{MB,2}$ $dv$) ratios are less than 0.2, which is
inconsistent with the assumptions we have made above. For these
cases, we adopted the mean derived excitation temperature of 15.3
$\pm$ 7.9 K for MYSOs and 7.7 $\pm$ 2.8 K for HII regions. The
column densities of N$_2$H$^+$ can be derived using Eq. (1) in Chen
et al. (2013). The derived parameters are listed in Table 4. The
derived optical depth of N$_2$H$^+$ (group 2) range from 0.10 to
2.3, with 79$\%$ below 1.0.

For HCO$^{+}$ and H$^{13}$CO$^+$ the assumptions inherent in the
LTE method are (1) the HCO$^+$ emission is optically thick, (2)
the molecules along the same LOS have a uniform excitation
temperature, and (3) the excitation temperature is the same for
the two isotopic species. With these assumptions, we calculate the
excitation temperature according to
\begin{equation}
T_{ex} = \frac{h v_0}{k} [ln (1 + \frac{h v_0/k }{T_{mb}/(1 -
e^{-\tau_v}) + J_v(T_{bg})})]^{-1}
\end{equation}
where $v_0$ is the rest frequency, $T_{bg}$ is the temperature of
the background radiation (2.73 K) and
\begin{equation}
J_v (T) = \frac{h v_0}{k} \frac{1}{e^{h v_0/k T} - 1}
\end{equation}
The optical depth of H$^{13}$CO$^+$ can be derived by
\begin{equation}
\tau = -ln [1- \frac{T_{mb}(H^{13}CO^+)}{J_v(T_{ex}) - J_v(T_{bg})}]
\end{equation}
Thus the column density of H$^{13}$CO$^+$ could also be determined
using Eq. (1) in Chen et al. (2013). The derived parameters are
listed in Table 4. It can be noted that H$^{13}$CO$^+$ (1-0) is
optically thin in all our sources, ranging from 0.18 to 0.67.
Assuming the [HCO$^+$]/[H$^{13}$CO$^+$] abundance ratio is 50
(e.g. Purcell et al. 2006), the optical depths of HCO$^+$ (1-0)
should be in the range of 9.0 to 33.5, consistent with our
assumptions. We also assumed H$^{13}$CO$^+$ (1-0) shares a common
excitation temperature. Using Eq. (7) of Wienen et al. (2012), we
recalculated the brightness temperature of H$^{13}$CO$^+$ (1-0).
The derived values are nearly the same as the observed ones,
suggesting beam dilution may be not a serious problem in our
analysis.

The optical depth of C$_2$H can be calculated through the intensity
ratio of its hyperfine components. Under the assumption of optically
thin, the intensity ratio of C$_2$H ($N=1-0, J=3/2-1/2, F=2-1$) and
C$_2$H ($N=1-0, J=3/2-1/2, F=1-0$) should be 2.0 (Tucker et al.
1974). Thus, the opacity of the brightest component ($F=2-1$) is
given by
\begin{equation}
\frac{1 - e^{-0.5\tau}}{1 - e^{-\tau}} =
\frac{T_{mb}(F=1-0)}{T_{mb}(F=2-1)}
\end{equation}
The excitation temperature of C$_2$H could not be derived directly.
We assume that its $T_{ex}$ is equal to that of N$_2$H$^+$, as
$\Delta V (C_2H)$ has a good correlation with $\Delta V (N_2H^+)$
(see section 3.1). Although C$_2$H is a linear molecular, its
rotational energy levels are described by the rotational quantum
number $N$ instead of $J$ (Sanhueza et al. 2012). Thus we used Eq.
(1) and table 7 in Sanhueza et al. (2012) to calculate the column
density of C$_2$H. The derived parameters are also listed in Table
4.

\begin{figure}
\psfig{file=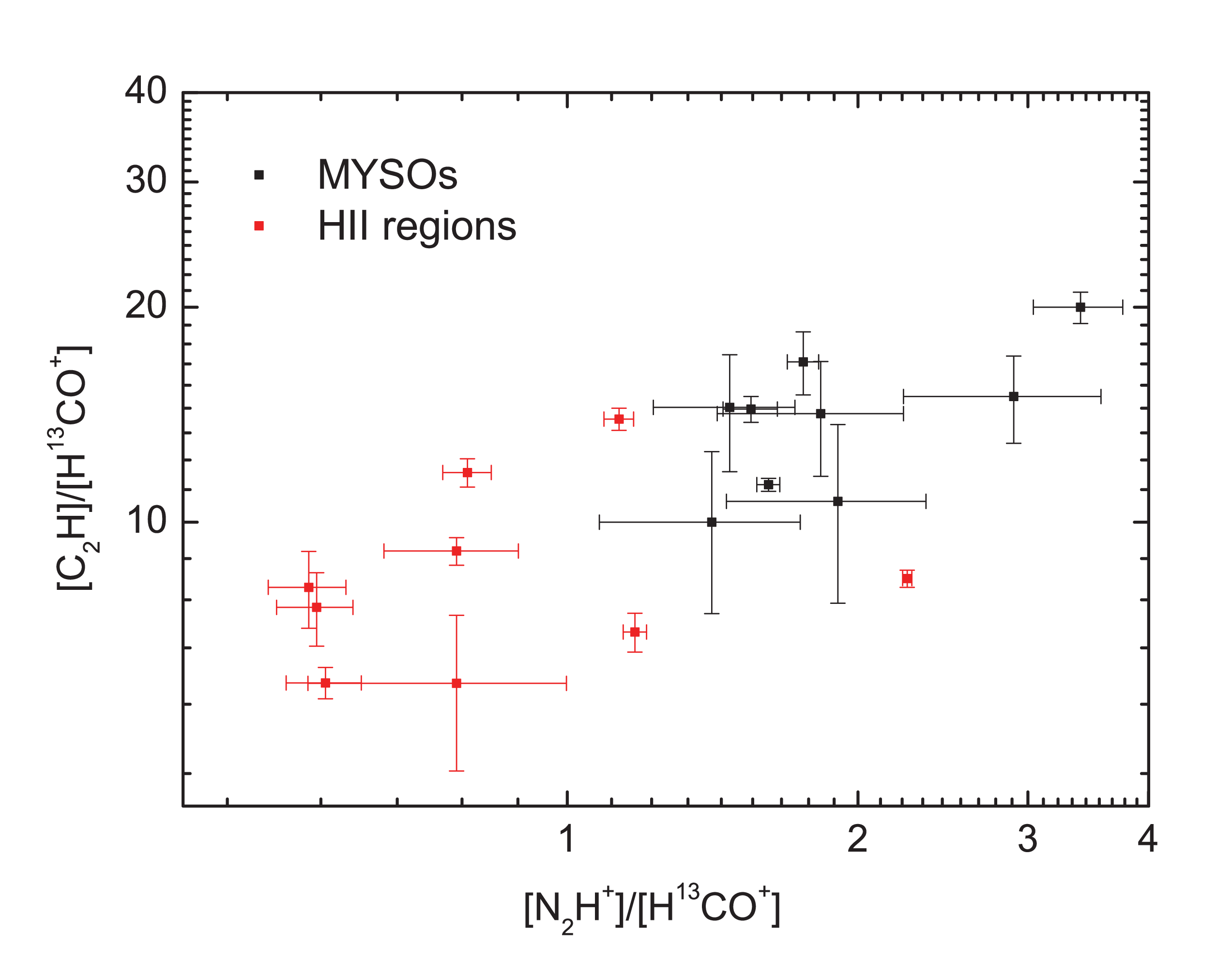,width=2.6in,height=2.0in} \caption{Relative
abundance rations of [N$_2$H$^+$]/[H$^{13}$CO$^+$] vs.
[C$_2$H]/[H$^{13}$CO$^+$].}
\end{figure}

\section{discussions}
As is mentioned above, N$_2$H$^+$ is widely detected in low- and
high-mass prestellar and protostellar cores, owing to its resistance
of depletion at low temperature and high densities. In star
formation regions, it is mainly destroyed  by CO and through
dissociative recombination with electrons produced by UV photons
from the central stars (e.g. Yu et al. 2015). In the chemical models
of low mass star formation, a relative enhancement of N$_2$H$^+$
abundance is expected in the cold prestellar phase, as CO is thought
to be depleted in starless cores (e.g. Lee et al. 2003; Bergin \&
Tafalla 2007). When the central star evolves, the gas gets warm and
CO should be evaporated from the dust grains if the dust temperature
exceeds about 20 K (Tobin et al. 2013). Thus the N$_2$H$^+$
abundance should decrease as a function of evolutionary stage. Our
column density of N$_2$H$^+$ ranges from 9.0 $\times$ 10$^{12}$ to
1.8 $\times$ 10$^{14}$ cm$^{-2}$, with an average of 6.3 $\times$
10$^{13}$ cm$^{-2}$. This value is consistent with that derived by
Miettinen (2014) and is also very similar to that of Sakai et al.
(2008). The top panel of Figure 5 shows column densities of
N$_2$H$^+$ versus those of H$^{13}$CO$^+$. It is apparent that
column densities of N$_2$H$^+$ in HII regions tend to be smaller
than those in MYSOs. The H$^{13}$CO$^+$ column density could reflect
the total amount of H$_2$ density in a clump as its abundance does
not vary so much with time according to the chemical models (e.g.
Bergin et al. 1997; Nomura $\&$ Millar 2004). HII regions are more
evolved than MYSOs and should have accreted more material, which
means the N$_2$H$^+$ column density should also increases. One good
reason for this is the depletion of N$_2$H$^+$ in the late stages of
massive star formation, probably caused by the formation of HII
regions inside.

C$_2$H was first detected by Tucker et al. (1974) in interstellar
clouds. However, it has not been systematically studied in massive
star formation regions. Only a few observations have been carried
out to study chemical evolution of C$_2$H in massive star
formation regions. C$_2$H could be formed through the
photodissociation of acetylene (C$_2$H$_2$): C$_2$H$_2$ + $h v$
$\rightarrow $ C$_2$H + H (Fuente et al. 1993). Thus it is
regarded to be a photodissociation region tracer. However,Beuther
et al. (2008) suggest C$_2$H may preferentially exist in the outer
part of dense gas clumps, as they found that the distribution of
C$_2$H shows a hole in the hot core phase. Given the large beam of
Mopra, we cannot know the detail spatial distributions of C$_2$H
in our EGO clumps. Our study shows that the C$_2$H velocity width
is as narrow as the HC$_3$N velocity width in the stage of MYSO.
However, in HII regions the C$_2$H velocity width tends to be
narrower than that of HC$_3$N, indicating the C$_2$H emissions do
not come from the stirred up gas in the center of the clumps. We
could also note that the best $\Delta$V correlation with C$_2$H is
N$_2$H$^+$, indicating their emissions might originate from the
same region. Li et al. (2012) observed 27 massive star formation
regions with water masers. They found the C$_2$H optical depth
declines when molecular clouds evolve to a later stage, suggesting
that C$_2$H might be used as a ``chemical clock" for molecular
clouds. In our sample, the average optical depth of C$_2$H is 0.88
in MYSOs and decreases to 0.48 in HII regions, consistent with
their results. Figure 5 shows column densities of C$_2$H versus
those of H$^{13}$CO$^+$. It seems like that the C$_2$H column
densities in HII regions also tend to be smaller than those in
MYSOs. Our result is consistent with the work of Beuther et al.
(2008) and Miettinen (2014), who found that both the mean and
median values of the C$_2$H abundance are higher towards IR-dark
clumps than towards IR-bright clumps. We guess in the hot dense
core, carbon-chain molecules like C$_2$H may rapidly transformed
to other molecules, however, most of the carbon atoms in the
central core are converted into CO, leading the production of
C$_2$H to be less effective. Since our limited data, more studies
should be done in the near future to check out this speculation.
Figure 6 shows the [N$_2$H$^+$]/[H$^{13}$CO$^+$] and
[C$_2$H]/[H$^{13}$CO$^+$] relative abundance ratios. The
difference between MYSOs and HII regions is remarkable. Our study
suggests depletions of N$_2$H$^+$ and C$_2$H in the late stages of
massive star formation, probably caused by the formation of HII
regions inside. N$_2$H$^+$ and C$_2$H might be used as ``chemical
clocks" for massive star formation by comparing with other
molecules such as H$^{13}$CO$^+$ and HC$_3$N.

\section{summary}
We present a molecular line study toward 31 EGO clumps in the
southern sky using data from the MALT90. We divide our sample into
two groups: MYSOs and HII regions, according to previous
multiwavelength observations. We have found that the velocity widths
of N$_2$H$^+$, C$_2$H, H$^{13}$CO$^+$ and HC$_3$N are comparable to
each other in MYSOs. However, in HII regions the velocity widths of
N$_2$H$^+$ and C$_2$H tend to be narrower than those of
H$^{13}$CO$^+$ and HC$_3$N. These results support that N$_2$H$^+$
and C$_2$H emissions mainly come from the gas inside quiescent
clumps. We also found column densities of N$_2$H$^+$ and C$_2$H
decrease from MYSOs to HII regions. In addition, the
[N$_2$H$^+$]/[H$^{13}$CO$^+$] and [C$_2$H]/[H$^{13}$CO$^+$]
abundance ratios also decrease with the evolutionary stage of the
EGO clumps. These results suggest depletions of N$_2$H$^+$ and
C$_2$H in the late stages of massive star formation, probably caused
by the formation of HII regions inside. N$_2$H$^+$ and C$_2$H might
be used as ``chemical clocks" for massive star formation by
comparing with other molecules such as H$^{13}$CO$^+$ and HC$_3$N.

\section*{ACKNOWLEDGEMENTS}
We thank the anonymous referee for constructive suggestions. This
paper made use of information from the Red MSX Source survey
database http://rms.leeds.ac.uk/cgi-bin/public/RMS$_{-}$DATABASE.cgi
and the ATLASGAL Database Server
http://atlasgal.mpifr-bonn.mpg.de/cgi-bin/ATLASGAL$_-$DATABASE.cgi.
The Red MSX Source survey was constructed with support from the
Science and Technology Facilities Council of the UK. The ATLASGAL
project is a collaboration between the Max-Planck-Gesellschaft, the
European Southern Observatory (ESO) and the Universidad de Chile.
This research made use of data products from the Millimetre
Astronomy Legacy Team 90 GHz (MALT90) survey. The Mopra telescope is
part of the Australia Telescope and is funded by the Commonwealth of
Australia for operation as National Facility managed by CSIRO. This
paper is supported by National Key Basic Research Program of China (
973 Program ) 2015CB857100.


\end{document}